\documentclass[dvipsnames]{article}

\usepackage[normalem]{ulem}
\usepackage{a4wide}

\usepackage{amsmath,amsfonts,amsthm,amssymb,bbm}
\usepackage[justification=justified, singlelinecheck=false]{caption}
\usepackage{graphicx}
\usepackage{enumitem}

\usepackage{tikz}
\usepackage{algorithmicx}
\usepackage{algorithm}
\usepackage{algpseudocode}

\usepackage[scaled=.90]{helvet}
\usepackage{courier}
\usepackage{ae}
\usepackage[T1]{fontenc}
\usepackage{subfigure}
\usepackage{graphicx,color}
\usepackage{xcolor}

\usepackage[T1]{fontenc}
\usepackage{here} 
\usepackage[colorlinks=false, pdfborder={0 0 0}]{hyperref}

\usepackage{xpatch}
\makeatletter
\xpatchcmd{\@thm}{\thm@headpunct{.}}{\thm@headpunct{}}{}{}
\makeatother

\newtheorem{remark}{Remark}

\theoremstyle{definition}

\begin{document}

\title{Inference for the  
stochastic FitzHugh-Nagumo model\\ from real action potential data\\ via approximate Bayesian computation}

\author{Adeline Samson\footnotemark[1]\thanks{Laboratoire Jean Kuntzmann, University Grenoble Alpes (Adeline.Leclercq-Samson@univ-grenoble-alpes.fr)}, Massimiliano Tamborrino\footnotemark[2]\thanks{Department of Statistics, University of Warwick (massimiliano.tamborrino@warwick.ac.uk)}, Irene Tubikanec\footnotemark[3]\thanks{Institute of Applied Statistics, Johannes Kepler University Linz (irene.tubikanec@jku.at)}\\}
\date{}
\maketitle	
	
\thispagestyle{empty}


\section*{Abstract}
The stochastic FitzHugh-Nagumo (FHN) model is a two-dimensional nonlinear  
stochastic differential equation with additive degenerate noise, whose first component, the only one observed, 
describes the membrane voltage evolution of a  single neuron. Due to its low-dimensionality, its analytical and numerical tractability and its neuronal interpretation, it has been used as a case study to test the performance of different statistical methods in estimating the underlying model parameters. Existing methods, however, often require complete observations, non-degeneracy of the noise or a complex architecture (e.g., to estimate the transition density of the process, \lq\lq recovering\rq\rq\ the unobserved second component) and they may not (satisfactorily) estimate all model parameters simultaneously. Moreover, these studies lack real data applications for the stochastic FHN model. 
The proposed method tackles all challenges (non-globally Lipschitz drift, non-explicit solution, lack of available transition density, degeneracy of the noise and partial observations). It is an intuitive and easy-to-implement 
sequential Monte Carlo approximate Bayesian computation algorithm, which relies on a recent computationally efficient and structure-preserving numerical splitting scheme for synthetic data generation and on summary statistics exploiting the structural properties of the process. All model parameters are successfully estimated from simulated data and, more remarkably, real action potential~data of rats. 
The presented novel real-data fit may broaden the scope and credibility of this classic and widely used neuronal model. 

\subsubsection*{Keywords} FitzHugh-Nagumo model, Action potential data, Hypoellipticity, 
Splitting numerical methods, Bayesian inference, Simulation-based inference, Sequential Monte Carlo approximate Bayesian computation.

\subsubsection*{AMS subject classifications} 60H10, 60H35, 65C30

\subsubsection*{Acknowledgements} 

A part of this paper was written while I.T. was member of the Department of Statistics, University of Klagenfurt, 9020 Klagenfurt, Austria. She is thankful for the support and the high-performance computing infrastructure provided by that university. 

\vspace{0.5cm}


\section{Introduction}
\label{sec:1_FHN}
The FitzHugh-Nagumo (FHN) model \cite{FitzHugh1961,Nagumo1962} is a well-known mathematical model, which is applied in the fields of theoretical and computational neuroscience and used to describe the pulses and bursts occurring in the membrane potential of a single neuron. The FHN model is a nonlinear second-order polynomial differential system bearing two states representing the membrane potential $V$ and a recovery variable $U$, respectively, with the latter comprising all ion channel related processes into one state. This system is less complex than the Hodgkin-Huxley \cite{Hodgkin1952} or the Morris-Lecar \cite{Morris1981} model, since it does not use physical parameters like ion conductances. Stochastic versions of these neuronal models have been proposed to describe various sources of randomness. In this paper, we focus on the case where the noise directly affects the second component (the variable $U$ of the FHN model), modelling the randomness of the conductance dynamics (see e.g. \cite{Bonaccorsi2008,Leon2018}). This leads to a hypoelliptic stochastic differential equation (SDE), i.e., an SDE with a degenerate noise structure but a smooth transition density (the membrane potential component $V$ is only indirectly impacted by noise). Measurements of the dynamics of a neuron, for example voltage-clamp experiments, provide observations of the membrane potential $V$ at discrete times, while the recovery variable $U$ is not observed. Such partially observed SDE is characterised by four parameters determining its behaviour, and it has been proved to be geometrically ergodic~in~\cite{Leon2018}.   

Our goal is to estimate all four parameters of the stochastic FHN model in a Bayesian setting. To achieve this, we need to face and tackle several challenges. First, the solution and the transition density of the SDE are not available in an explicit closed-form, implying that numerical methods are needed to approximate them. Second, the SDE has a degenerate noise structure preventing using classical numerical schemes, such as the Euler-Maruyama discretisation, for likelihood approximation, as these schemes do not share this property  \cite{Buckwar2022,Ditlevsen2019,Melnykova2020,Pokern2009}. Third, the drift of the SDE is non-globally Lipschitz, and thus standard It\^o-Taylor numerical methods do not converge to the true solution (in the strong mean-square sense) \cite{Hutzenthaler2011}. This problem may be overcome by using tamed \cite{Hutzenthaler2012_2,Sabanis2013,Tretyakov2012,Zhang2017} or truncated \cite{Mao2017,Hutzenthaler2012,Mao2015,Mao2016} variants. However, this type of methods still fails to be structure-preserving, meaning that they do not capture important properties of the model, such as its hypoellipticity (as discussed before), geometric ergodicity or oscillatory dynamics (see \cite{Buckwar2022} and the references therein). Fourth, only partial observations of the system are available. This is difficult because the coordinate $V$ alone is not Markovian, only the couple $(V,U)$ shares this property. 

In the frequentist framework, several estimation methods applicable to the stochastic FHN model have been proposed. They use adapted numerical schemes and are based on estimation contrasts or optimal control theory. Among others, we can cite \cite{Iguchi2023,Melnykova2020,Pokern2009} in the case of complete observations and 
\cite{Clairon2022,Ditlevsen2019,Gloter2021,Iguchi2022,Samson2012} in the case of partial observations. 
In the Bayesian setting, we refer to \cite{Grahametal2022,vanderMeulen2020}, where specific filtering algorithms,  based on either the Taylor 1.5 scheme or the Euler-Maruyama scheme, are applied to derive the marginal posterior distributions of the FHN model parameters. 

In this paper, we focus on the approximate Bayesian computation (ABC) method, see \cite{Marin2012, sisson2018handbook} for a review. ABC is a simulation-based inference approach for Bayesian parameter estimation in complex mathematical models with unknown or intractable likelihood functions. In its basic acceptance-rejection version, it derives an approximate posterior distribution for the parameters of interest in three simple steps. First, a parameter value is sampled from a proposal distribution (e.g., the prior, for the simple acceptance-rejection ABC algorithm). Second, synthetic data are simulated from the model using the sampled parameter value. Third, such value is kept as a realisation from the approximate posterior distribution if the distance between some summary statistics of the reference data and the simulated data is smaller than some predefined tolerance level. This procedure is then repeated several times until, e.g., a certain number of sampled parameters are accepted. Among others, ABC has been proposed for inference in time series models \cite{Drovandi2016, Jasra2015}, state space models \cite{Martin2019, Tancredi2019} and SDE models, see e.g. \cite{Buckwar2019,Ditlevsenetal2023, Jovanovskietal2024, Kypraios2017, Maybank2017, Picchini2014}. ABC has, however, not yet been applied to the stochastic FHN model, and, more generally, to hypoelliptic SDEs with locally Lipschitz drift coefficients, for which a convergent structure-preserving numerical scheme has been recently proposed in \cite{Buckwar2022}, as discussed below. 

As ABC relies on running extensive simulations from the model under different parameter values (step two), the quality of its results depends strongly on the choice of the numerical scheme used for synthetic data generation. 
Due to its simple implementation and  properties (e.g.  Gaussian transition densities), the standard Euler-Maruyama scheme has been used for data-simulation from any SDE model (see, e.g., \cite{Picchini2014,PicchiniForman2016,PicchiniSamson2018,Sun2015}), including the stochastic FHN. However, in \cite{Buckwar2019} it has been recently shown that such scheme may lead to drastic fails in the estimation, 
due to its non-structure-preserving nature. Instead, adopting a structure-preserving numerical scheme led 
to successful inference of the underlying parameters (from partially observed and hypoelliptic  Hamiltonian-type SDEs, including oscillatory neuronal models). Following this, here we use a numerical splitting scheme for the data-simulation within ABC. This method was recently proposed in \cite{Buckwar2022} and shown to be both \textit{mean-square convergent} (of order $1$) and \textit{structure-preserving}, i.e., it retains the hypoellipticity, geometric ergodicity and oscillatory dynamics (amplitudes, frequencies and phases) of the stochastic FHN model, even for large time-discretisation steps. Hence, it is
not only reliable, but also computationally efficient, which is particularly important when used within a simulation-based algorithm requiring millions of model simulations.

Besides the chosen numerical scheme, the performance of ABC also depends heavily on the choice of suitable summary statistics for the investigated data. When applying ABC to SDE models, it is particularly important to choose summary statistics that are sensitive to small parameter changes and robust to the intrinsic stochasticity of the model. Inspired by \cite{Buckwar2019},  we propose to use statistics that are invariant for repeated simulations under the same parameter setting, namely the invariant density and spectral density, which are both based on the geometric ergodicity property of the stochastic FHN model. Moreover, we show how such choice leads to more accurate approximate posterior densities than those obtained using standard canonical summaries (e.g., mean, variance and auto-correlation).

Differently from \cite{Buckwar2019}, we do not embed here the structure-based summaries and structure-preserving numerical splitting scheme within a basic acceptance-rejection ABC algorithm. Indeed, such algorithm 
is very computationally wasteful, since parameter proposals are sampled from the prior distribution, which typically covers regions far away from the desired posterior distribution mass.  Thus, many parameter candidates are rejected, after having spent computational time and effort to simulate the corresponding synthetic data, the summaries and the distance measure. To improve this, several advanced ABC algorithms have been introduced over the years, such as sequential Monte Carlo (SMC) ABC (the gold standard state-of-art algorithm), Markov Chain Monte Carlo (MCMC) ABC, or sequential importance sampling (SIS) ABC, see \cite{sisson2018handbook} for a review. While such sequential ABC approaches have been applied to SDE models (see, e.g., \cite{Jovanovskietal2024,Picchini2014}), these works do typically not consider structure-based summaries and structure-preserving simulation methods. In this paper, we employ the structure-based ingredients within an efficient SMC-ABC scheme, yielding what we call \textit{structure-based and preserving (SBP) SMC-ABC method}.

Our SBP SMC-ABC method succeeds in estimating all parameters of the stochastic FHN model from simulated data, even for a relatively small number of observations and a short observation time horizon. More interestingly, we also apply our method to real data consisting of membrane voltages recorded in an adult female Sprague Dawley rat \cite{METCALFE2020}. We analyse four recordings, two obtained under a resting condition and the other two under a stimulation setting. We show the coherence of the estimated posterior distributions with respect to these experimental setups. The fitted models are then validated through goodness-of-fit plots obtained by simulations under the estimated model. 

While there exist only very few results fitting (or using) the  FHN model to (for) real data applications, see e.g. \cite{Doruk2019} for a study on neural spike time data and \cite{LOPEZ2017} for an investigation of vibration bearing data, this is, to the best of our knowledge, the first work on the considered stochastic FHN model investigating and fitting some real action potential data. This is a novel result which potentially broadens the models' scope and credibility. In addition, note that the proposed method is not limited to the stochastic FHN model, but it can be applied to any other ergodic SDE for which a computationally tractable structure-preserving numerical method is available. 

\newpage

The paper is organised as follows. In Section \ref{sec:2_FHN}, we introduce the stochastic FHN model and discuss its mathematical characteristics and structural properties. In Section \ref{Section3}, we introduce the SBP SMC-ABC method, detailing the algorithm and all required key ingredients. In Section~\ref{sec:4_FHN}, we use this method to estimate the parameters of the stochastic FHN model from simulated data, under different observation settings. Section \ref{sec:5_FHN} details the results obtained on the real action potential data. Conclusions are reported in Section \ref{sec:6_FHN} and details on the algorithm and results are presented in the appendix.


\vspace{-0.3cm} 
\section{Stochastic FHN model}
\label{sec:2_FHN}
\vspace{-0.2cm} 

The stochastic FHN model \cite{Buckwar2022,Leon2018} is a well-established prototype model of a single neuron (nerve cell) and dates back to the work of FitzHugh (1961) \cite{FitzHugh1961} and Nagumo (1962)~\cite{Nagumo1962}. It describes the generation of spikes (also called action potentials) at the intracellular level. Noise can be included either on the membrane potential or on the conductance dynamics of the model. Here, we consider the randomness of opening and closing ion channels at the surface of the neuron membrane through noise acting on the conductance dynamics \cite{Pakdaman2010}. The resulting process can be obtained as solution to the following two-dimensional nonlinear SDE
\begin{equation}\label{FHN}
	d \underbrace{\begin{pmatrix}
			V(t) \\
			U(t) 
	\end{pmatrix}}_{:=X(t)}
	=
	\underbrace{\begin{pmatrix}
			\frac{1}{\epsilon}\Bigl(V(t)-V^3(t)-U(t)\Bigr) \\
			\gamma V(t)-U(t)+\beta
	\end{pmatrix}}_{:=F(X(t))} dt \ + \
	\underbrace{\begin{pmatrix}
			0 \\ \sigma
	\end{pmatrix}}_{:=\Sigma} dW(t), \quad X(0)=X_0, \quad t\in[0,T],
\end{equation}
where $(W(t))_{t\in [0,T]}$ is a standard Wiener process and the initial value $X_0=(V_0,U_0)^\top$ is a $\mathbb{R}^2$-valued random variable with bounded moments that is independent of $(W(t))_{t\in [0,T]}$. This is a  
common formulation of the model, where no noise enters directly into the $V$-component of SDE~\eqref{FHN} and only the $U$-component is directly perturbed by it. The considered stochastic FHN model contains four parameters 
\begin{equation}\label{eq:theta}
	\theta=(\epsilon,\gamma,\beta,\sigma),
\end{equation}
where $\beta,\gamma>0$ describe the position and duration of an action potential, respectively, $\epsilon>0$ is a time scale separation parameter and $\sigma>0$ defines the noise intensity. 
The $V$-component of system \eqref{FHN} models the evolution of the membrane voltage of the neuron, which can be measured with suitable recording techniques (cf. Section \ref{sec:5_FHN}),  while the $U$-component is an unobserved recovery variable modelling the ion channel kinetics.

\paragraph{Mathematical characteristics and structural properties}

The drift $F(X(t))$ of the FHN model~\eqref{FHN} is not globally Lipschitz continuous. However, as it satisfies a  \textit{one-sided Lipschitz} condition \cite{Brehier2024,Buckwar2022}, the SDE has a unique strong (unknown) solution $X(t)=(V(t),U(t))^\top$, $t\in [0,T]$, which is regular in the sense of \cite{Khasminskii2011}, that is, it is defined on the whole time interval $[0,T]$ and paths of the process do not blow up to infinity in finite time. Moreover, such solution is a Markov process with (unknown) transition probability 
\vspace{-0.2cm} 
\begin{equation}\label{eq:trans_prob}
	P_{t}(\mathcal{A},x):=\mathbb{P}\left( X(t) \in \mathcal{A} | X(0)=x \right), \quad \mathcal{A} \in \mathcal{B}(\mathbb{R}^2),
\end{equation}
where $\mathcal{B}(\mathbb{R}^2)$ denotes the Borel sigma-algebra on $\mathbb{R}^2$. 

Furthermore, the process $(X(t))_{t\in[0,T]}$ possesses two known structural properties. First, it is \textit{hypoelliptic} \cite{Ditlevsen2019,Leon2018}. This means that the transition distribution, defined via \eqref{eq:trans_prob}, has a smooth density, even though  the squared diffusion term 
\begin{equation*}\label{eq:Sigma_squared}
	\Sigma\Sigma^\top=\begin{pmatrix}
		0 & 0\\
		0 & \sigma^2
	\end{pmatrix}
\end{equation*}
is not of full rank, due to the degenerate noise structure of the FHN model \eqref{FHN}. This property holds because the $U$-component (the model coordinate which is directly perturbed by the noise) enters into the first component of the drift $F(X(t))$.

Second, it is \textit{geometrically ergodic} \cite{Ableidinger2017,Buckwar2022,Leon2018,Mattingly2002}, i.e., the distribution of the solution  converges exponentially fast  to a unique invariant distribution~$\eta$ for any initial value $X_0$, with $\eta$ satisfying
\begin{equation*}
	\eta(\mathcal{A})=\int\limits_{\mathbb{R}^2}P_t(\mathcal{A},x)\eta(dx), \quad \forall \ \mathcal{A} \in \mathcal{B}(\mathbb{R}^2), \ t \in [0,T].
\end{equation*}
In particular, if $X_0$ has distribution $\eta$, then $X(t)$ has distribution $\eta$ for all $t \in [0,T]$. The existence of the invariant distribution $\eta$ follows from the fact that there exists a Lyapunov function  
for system~\eqref{FHN}. For further details regarding the characteristics and structural properties of the stochastic FHN model, the reader is referred to \cite{Buckwar2022,Leon2018} and the references therein.

\paragraph{Statistical setting}

The stochastic FHN model \eqref{FHN} is  only \textit{partially observed} through discrete-time measurements of its first coordinate, the $V$-component. The measurements are made at equidistant times $0=t_0<t_1 < \ldots < t_{n-1} < t_n=T_{\textrm{obs}}$, with $t_i=i\Delta_{\textrm{obs}}$,  $\Delta_\textrm{obs}=T_{\textrm{obs}}/n$,  $i=0,\ldots,n$, where $T_{\textrm{obs}}$ and $\Delta_{\textrm{obs}}$ denote the observation time horizon and time step, respectively. Here, the resulting observation vector is denoted by $y=(V(t_i))_{i=0}^{n}$. The goal of this paper is to derive an approximate posterior density (via the ABC method) for $\theta$ \eqref{eq:theta}, given the available observations $y$. 


\vspace{-0.3cm}
\section{Structure-based and preserving SMC-ABC method}\label{Section3}
\vspace{-0.2cm}

In this section, we introduce the proposed ABC methodology for parameter estimation of the stochastic FHN model \eqref{FHN}. In Section \ref{sec:3:SMC-ABC-alg}, we recall the gold-standard SMC-ABC algorithm. In Section \ref{sec:3:ABCingredients}, we detail our proposed key ingredients of this algorithm, yielding the SBP SMC-ABC method. In Section \ref{sec:3:impl.det.}, we provide implementation details.


\subsection{SMC-ABC algorithm}
\label{sec:3:SMC-ABC-alg}

Let $\pi(\theta)$ denote the prior density of an unknown parameter $\theta$, $p(y|\theta)$ the (unavailable) likelihood function of observations $y$ and $\pi(\theta|y)\propto p(y|\theta)\pi(\theta)$ the desired posterior density. We now recall the basic acceptance-rejection ABC algorithm presented in the introduction, to introduce the relevant notation. Such scheme consists of three steps: (a) sample $\theta'$ from the prior $\pi(\theta)$; (b) conditioned on $\theta'$, simulate a synthetic dataset $\tilde y_{\theta'}$ from the model; (c) keep the sampled value $\theta'$ if the distance $d(\cdot, \cdot)$ between the vector of summary statistics $s(\cdot)$ of the observed and simulated data is smaller than a user-defined tolerance level $\delta$. This leads to an approximate posterior given~by 
\begin{equation*}
	\pi(\theta|y)\approx \pi_\textrm{ABC}^\delta(\theta|s(y))\propto \int\mathbbm{1}_{\{d(s(y),s(\tilde y_{\theta'}))<\delta\}}\pi(\theta)p(s(\tilde y_{\theta'})|\theta)ds.
\end{equation*} 
As this ABC scheme is highly computationally inefficient, sequential ABC schemes have then been proposed to tackle this, with SMC-ABC being the most prominent one.  The idea of SMC-ABC is to construct a sequence of $r$ intermediate ABC posteriors using proposal samplers (also known as perturbation kernels), which are 
based on the $N$ kept sampled values (called particles) at the previous iterations, and are thus continuously updated across iterations. 

The first iteration of SMC-ABC coincides with acceptance-rejection ABC with a tolerance level $\delta_1$, after which all kept particles $(\theta_1^{(1)}, \ldots, \theta_1^{(N)})$ are given an equal weight $w_{1}^{(j)}=1/N$, $j=1,\ldots, N$, and a smaller threshold $\delta_2$ is proposed. At each iteration $r>1$, a value $\theta_j$ is initially sampled from the set of kept particles $(\theta_{r-1}^{(j)})_{j=1}^N$ at
iteration $r-1$ according to the corresponding weights $(\omega_{r-1}^{(j)})_{j=1}^N$, and then perturbed via a proposal sampler $K_r(\cdot|\theta_j)$, yielding $\theta_j^*.$ Such value is then used to simulate new data $\tilde y_{\theta_j^*}$, and $\theta_j^*$ is accepted if $d(s(y),s(\tilde y_{\theta_j^*}))<\delta_r$. After $N$ particles are accepted, the unnormalised weights are updated as 
\begin{equation*}
	\tilde\omega_r^{(j)}=\pi(\theta_r^{(j)})/\sum_{l=1}^N \omega_{r-1}^{(l)}k_r(\theta_r^{(j)}|\theta_{r-1}^{(l)}), \quad j=1,\ldots,N,
\end{equation*} 
and then normalised via $\omega_r^{(j)}=\tilde\omega_r^{(j)}/\sum_{l=1}^N\tilde\omega_r^{(l)}$. Here, $k_r(a|b)$ denotes the density of the proposal sampler $K_r( \cdot | b)$ evaluated at $a$. 

The algorithm terminates when a stopping criterion is met, e.g. $\delta_r$ becomes smaller than a desired final threshold $\delta_\textrm{final}$, a certain number of iterations is reached, or the acceptance rate of a particle becomes lower than 1.5\% for two consecutive iterations \cite{DelMoral2012}. Here, the algorithm terminates when a computational budget $\textrm{Nsim}_\textrm{max} \gg 0$, consisting of the desired total number of synthetic model simulations, is reached. The corresponding procedure is detailed in Algorithm \ref{alg:SMC_SBP_ABC}. Note that, the computational budget $\textrm{Nsim}_\textrm{max}$ is exceeded in the last iteration, so the used budget is larger than the specified one.

\begin{algorithm}
	\caption{SMC-ABC (with computational budget as stopping criterion)
		\ \\ \textbf{Input:} Observed dataset $y$, summary statistics $s(\cdot)$, prior distribution $\pi(\theta)$, perturbation kernel~$K_r$, computational budget $\textrm{Nsim}_\textrm{max}$ (i.e. maximum number of simulated datasets), 
		number of kept samples in each iteration $N$ and starting threshold $\delta_1$  
		\ \\
		\textbf{Output:} Weighted samples from the SMC-ABC posterior distribution
	}\label{alg:SMC_SBP_ABC}
	\begin{algorithmic}[1]
		\State Set $r=1$ and $\textrm{Nsim}=0$.
		\For{$j=1:N$}
		\Repeat
		\State Sample $\theta_j$ from the prior $\pi(\theta)$.
		\State Conditioned on $\theta_j$,  
		simulate a synthetic dataset $\tilde{y}_{\theta_j}$ from the model,\newline and set   $\textrm{Nsim}=\textrm{Nsim}+1$ 
		\State Compute the summaries $s(\tilde{y}_{\theta_j})$. 
		\State Calculate the distance  $D_j=d\bigl(s(y),s(\tilde{y}_{\theta_j})\bigr).$ 
		\Until{$D_j<\delta_1$}
		\State Set $\theta_1^{(j)}=\theta_j$.
		\State Set $\tilde\omega_1^{(j)}=1$, $\omega_1^{(j)}=1/N$.
		\EndFor
		\While {$\textrm{Nsim}<\textrm{Nsim}_\textrm{max}$}
		\State Set $r=r+1$.
		\State Set $\delta_r< \delta_{r-1}$.
		\For{$j = 1:N$}
		\Repeat 
		\State Sample $\theta_j$ from the weighted set $\{(\theta_{r-1}^{(j)},\omega_{r-1}^{(j)})_{j=1}^N$\}.
		\State Perturb $\theta_j$ to obtain $\theta_j^{*}$ from $K_r(\cdot|\theta_j)$.
		\State If $\pi(\theta_j^{*})=0$, return to Line 17.
		\State Conditioned on $\theta_j^*$,  simulate a synthetic dataset $\tilde{y}_{\theta_j^*}$ from the model,\newline and set $\textrm{Nsim}=\textrm{Nsim}+1$
		\State Compute the summaries $s(\tilde{y}_{\theta_j^*})$. 
		\State Calculate the distance  $D_j=d\bigl(s(y),s(\tilde{y}_{\theta_j^*})\bigr)$. 
		\Until{$D_j<\delta_r$}
		\State Set $\theta_{r}^{(j)}=\theta_j^*$.
		\State Set  $\tilde{w}_r^{(j)}=\pi\left(\theta_{r}^{(j)}\right) / \sum\limits_{l=1}^{{N}} w_{r-1}^{(l)} k_r\left( \theta_{r}^{(j)} \Big| \theta_{r-1}^{(l)} \right)$.
		\EndFor
		\State Normalise the weights $w_r^{(j)}=\tilde{w}_r^{(j)}/ \sum\limits_{l=1}^{{N}} \tilde{w}_{r}^{(l)} $, for $j=1,\ldots,{N}$.
		\EndWhile
		\State 
		Return a set of weighted parameters $(\theta_r^{(1)},\omega_r^{(1)}),\ldots, (\theta_r^{(N)},\omega_r^{(N)})$.
	\end{algorithmic}
\end{algorithm}


\subsection{Ingredients for the SMC-ABC algorithm}
\label{sec:3:ABCingredients}

The SMC-ABC Algorithm \ref{alg:SMC_SBP_ABC} requires the choice of several key ingredients: the data simulation method, the summaries (along with a suitable distance measure), the prior distribution, the proposal sampler, the threshold levels and the stopping criterion (here, the allowed maximum number of synthetic data simulations). Among all, the first two play a crucial role, as we would like to apply an efficient numerical scheme capable of simulating from the model while preserving its key (structural) features, as well as to choose some summary statistics based on such properties. The chosen ingredients, detailed below, guarantee this, yielding the proposed SBP SMC-ABC method. 


\subsubsection{Structure-preserving synthetic data simulation with a splitting scheme}

A crucial step within each simulation-based statistical method is the generation of synthetic datasets from the model for a given parameter value (cf. Lines $5$ and $20$ of Algorithm \ref{alg:SMC_SBP_ABC}). As exact simulation schemes are not available for the stochastic FHN model \eqref{FHN} and standard It\^o-Taylor type methods are non-convergent and/or non-preserving, we will simulate data from SDE~\eqref{FHN} via the numerical splitting scheme recently proposed in \cite{Buckwar2022}.  

This method is based on the idea of re-writing the drift coefficient $F(X(t))$ of the stochastic FHN model \eqref{FHN} as the sum of a linear and a non-linear part, and then splitting the model into two exactly solvable subequations: a linear SDE, which shares the properties of hypoellipticity and geometric ergodicity, and a non-linear ordinary differential equation (ODE) capturing the non-globally Lipschitz polynomial term. The exact solutions of these two subequations are then derived and, in every iteration step, composed in a suitable way (here, according to the Strang approach, which considers ``half-steps'' of the ODE-solution). Due to the exact treatment of relevant and cleverly chosen subparts of the stochastic FHN model \eqref{FHN}, the resulting (Strang) splitting scheme preserves its structural properties (hypoellipticity, geometric ergodicity and oscillatory dynamics).  
The path generation procedure required in Algorithm \ref{alg:SMC_SBP_ABC} using this splitting scheme is briefly summarized in the following. For details regarding its construction and features, the interested reader is referred to \cite{Buckwar2022}.

For the model simulation, we consider an equidistant time grid $0 = t_0<t_1 <\ldots<t_{n-1}<t_n = T$, with $t_i = i\Delta$, $\Delta=T/n$, $i = 0,\ldots, n$. We set $\widetilde X(t_0) := X_0$ and denote by  $\widetilde X(t_i)$ the splitting approximation  of the solution
$X(t_i)$ of SDE \eqref{FHN} at time $t_i$. The considered splitting scheme relies on the linear SDE $dX_t=AX(t)+\Sigma dW(t)$, with
\begin{equation*}
	A=\begin{pmatrix}
		0 & -\frac{1}{\epsilon} \\
		\gamma & -1
	\end{pmatrix},
\end{equation*}
whose exact solution depends on the matrix exponential $E(t):=e^{At}$ arising from the drift coefficient and the covariance matrix
\begin{equation}
	C(t)=\int\limits_{0}^{t} E(t-s)\Sigma\Sigma^\top \left( E(t-s) \right)^\top ds.
\end{equation}
The resulting process constitutes a weakly-, critically-, or over-damped stochastic harmonic oscillator, depending on whether the value
\begin{equation}\label{eq:kappa}
	\kappa:=\frac{4\gamma}{\epsilon}-1
\end{equation}
is positive, zero, or negative, respectively, see e.g. Chapter 5 of \cite{Weiglhofer1999}. Requiring $\kappa>0$ (i.e., considering the weakly damped case), guarantees that the model produces suitable oscillatory dynamics. In this case, the matrix exponential and covariance matrix are given by
\begin{equation}\label{eq:MatrixExp}
	E(t):=e^{-\frac{t}{2}}\begin{pmatrix}
		\cos(\frac{1}{2}\sqrt{\kappa}t)+\frac{1}{\sqrt{\kappa}}\sin(\frac{1}{2}\sqrt{\kappa}t) & -\frac{2}{\epsilon\sqrt{\kappa}}\sin(\frac{1}{2}\sqrt{\kappa}t) \\
		\frac{2\gamma}{\sqrt{\kappa}}\sin(\frac{1}{2}\sqrt{\kappa}t) & \cos(\frac{1}{2}\sqrt{\kappa}t)-\frac{1}{\sqrt{\kappa}}\sin(\frac{1}{2}\sqrt{\kappa}t)
	\end{pmatrix},
\end{equation}
and 
\begin{equation}\label{eq:Cov}
	C(t):=\begin{pmatrix}c_{11}(t)&c_{12}(t)\\c_{12}(t)&c_{22}(t)\end{pmatrix},
\end{equation}
where  
\begin{eqnarray*}
	c_{11}(t)&=&\frac{ \sigma^2 e^{-t}}{2\epsilon\gamma\kappa} \biggl( -\frac{4\gamma }{\epsilon}  + \kappa e^t + \cos(\sqrt{\kappa}t) -\sqrt{\kappa} \sin(\sqrt{\kappa}t) \biggr), \\
	c_{12}(t)&=&\frac{\sigma^2 e^{-t}}{\kappa \epsilon} \biggl(\cos(\sqrt{\kappa}t) -1  \biggr), \\
	c_{22}(t)&=&\frac{\sigma^2 e^{-t}}{2\kappa} \biggl(  \cos(\sqrt{\kappa}t) +  \sqrt{\kappa} \sin(\sqrt{\kappa}t)-\frac{4\gamma}{\epsilon}+\kappa e^t  \biggr),
\end{eqnarray*}
respectively, for $t>0$.
Finally, we define the function that solves exactly the ODE subequation of the splitting framework
\[ h(x;t):=\begin{pmatrix}
	v/{\sqrt{e^{-\frac{2t}{\epsilon}}+v^2\left(1-e^{-\frac{2t}{\epsilon}}\right)}} \\
	\beta t + u
\end{pmatrix},
\]
for $t>0$, $x=(v,u)^\top \in \mathbb{R}^2$. A path of the FHN model \eqref{FHN} is then obtained via the Strang splitting scheme reported in Algorithm \ref{alg:Splitting}. Note that, this algorithm generates a trajectory for both coordinates $\widetilde X(t_i)=(\widetilde V(t_i),\widetilde U(t_i))^\top$,  $i=0,\ldots,n$, but only the first coordinate $\tilde y_\theta=(\widetilde V(t_i))_{i=0}^n$ is used in the inferential procedure.

\begin{algorithm}[t]
	\caption{Synthetic-data simulation of the FHN model \eqref{FHN} via Strang splitting \cite{Buckwar2022}
		\ \\ \textbf{Input:} Initial value $X_0$, step-size $\Delta$, number of time discretisation steps $n$, model parameters $\theta=(\epsilon,\gamma,\beta,\sigma)$ such that $\kappa$ \eqref{eq:kappa} is positive. 
		\ \\
		\textbf{Output:} Simulated path $(\widetilde X(t_i))_{i=0}^n$ of \eqref{FHN}, with $\widetilde y_\theta=(\widetilde V(t_i))_{i=0}^n$ used for inference}
	\label{alg:Splitting}
	\begin{algorithmic}[1]
		\State Compute the matrices $E(\Delta)$ \eqref{eq:MatrixExp} and $C(\Delta)$ \eqref{eq:Cov}. 
		\State Set $\widetilde{X}(t_0)=X_0$.
		\For{$i = 1:n$}
		\State Generate $\xi=(\xi_1,\xi_2)^\top$ from a bivariate normal distribution $\mathcal{N}\bigl( (0,0)^\top,
		C(\Delta)\bigr)$.
		\State Set $a=h(\widetilde{X}(t_{i-1});\Delta/2)$.  
		\State Set $b=E(\Delta)a+\xi$.  
		\State Set $\widetilde{X}(t_i)=h(b;\Delta/2)$.   
		\EndFor
		\State Return $\widetilde{X}(t_i)$,  
		for  $i=0,\ldots,n$ and use $\tilde y_\theta=(\widetilde V(t_i))_{i=0}^n$ for inference.
	\end{algorithmic}
\end{algorithm}

\begin{remark}\label{remark2}
	In our experiments, the prior will be chosen to meet the condition $\kappa>0$ in Algorithm~\ref{alg:Splitting}, see Section \ref{sectionprior}. However, this requirement can be relaxed, and we refer to \cite{Buckwar2022} for the corresponding expressions of $E(t)$ and $C(t)$ for $\kappa\leq 0$.
\end{remark}

\subsubsection{Structure-based data summaries}

\begin{figure}
	\begin{centering}
		\includegraphics[width=1.0\textwidth]{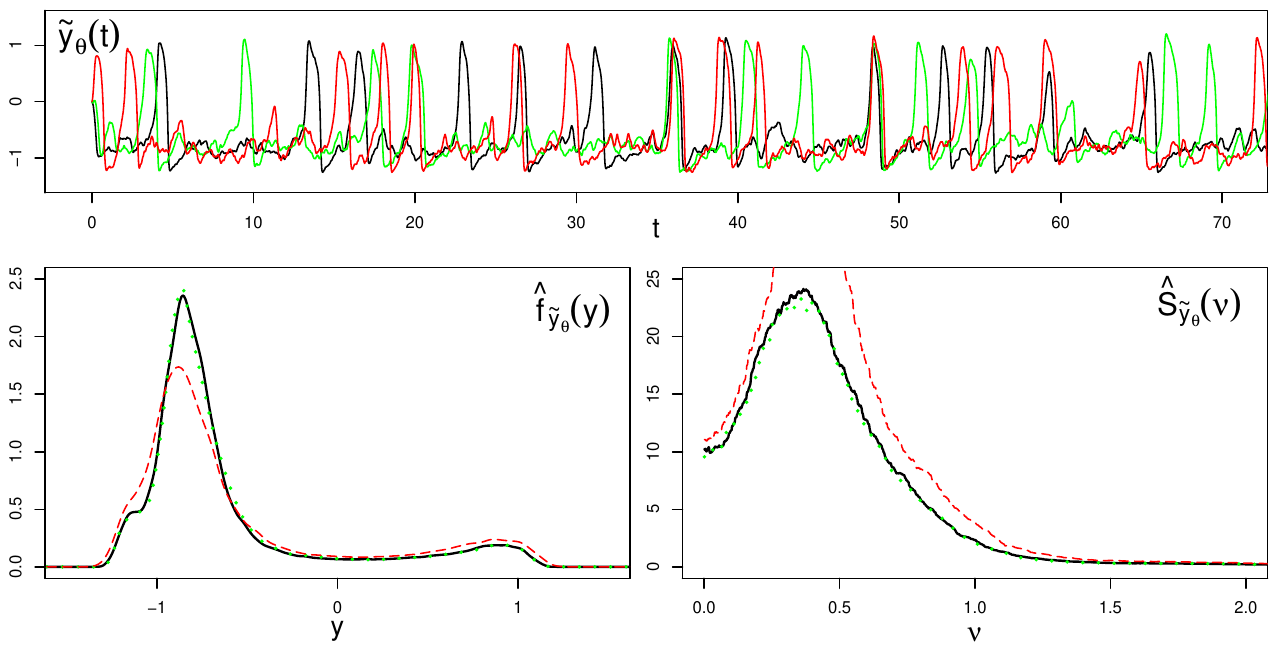}	
		\caption{Top panel: Three paths of the observed component $V(t)$ of SDE \eqref{FHN} generated with Algorithm \ref{alg:Splitting}. The parameter values are $\theta=(0.1,1.5,0.8,0.3)$ for the black and green paths and $\theta=(0.1,1.6,0.8,0.4)$ for the red path. Bottom panels: Corresponding estimated invariant density (left) and spectral density (right).}
		\label{fig:summaries}
	\end{centering}
\end{figure}

The summary statistics employed here (cf. Lines 6 and 21 of Algorithm \ref{alg:SMC_SBP_ABC}) are the structure-based summaries proposed in \cite{Buckwar2019} for geometrically ergodic SDEs. In particular, we consider the invariant density $f_\theta$ and the invariant spectral density $S_\theta$ given by 
\[
S_\theta(\nu)=\int_\mathbb{R} R_\theta(\tau)\exp(-i2\pi\nu\tau)d\tau,
\]
which is the Fourier transform of the auto-correlation function $R_\theta(\tau)$ with respect to the frequency~$\nu$. Here, $f_\theta$ and $S_\theta$ are estimated based on a given dataset $\tilde y_\theta$, via a standard kernel density estimator $\hat{f}_{\tilde y_\theta}$ \cite{Pons2011} and a smoothed periodogram estimator  $\hat{S}_{\tilde y_\theta}$ \cite{Quinn2014}, respectively, yielding the summaries
\begin{equation}\label{s}
	s(\tilde y_\theta):=(\hat{f}_{\tilde y_\theta},\hat{S}_{\tilde y_\theta}).
\end{equation}
Thanks to the ergodicity of the SDE and its preservation by the numerical splitting scheme in Algorithm~\ref{alg:Splitting}, these summaries have several desired features. First, they can be well estimated from a single path $\tilde y_\theta$ generated over a sufficiently long time horizon, instead of requiring a large number of repeated path simulations. Second, they are negligibly impacted by the (possibly unknown) initial value $X_0$. 
Moreover, as they are deterministic functions of the unknown parameter $\theta$, they fully characterise the asymptotic behaviour of the SDE (third), they are sensitive to small changes in $\theta$ (fourth) and robust 
to repeated simulations under the same parameter values (fifth).  These characteristics are illustrated in Figure \ref{fig:summaries}, where the top panel reports two paths obtained under the same value of $\theta$ 
(green and black trajectories) and a third path obtained under a slightly different value for $\theta$ (red trajectory). The corresponding summaries, derived from these datasets, are shown in the bottom panels. While judging on the similarities/differences between the trajectories seems difficult, this becomes straightforward when looking at their summaries, which are indistinguishable only when estimated from data generated under the same parameter~value. 

\begin{remark}\label{rem:s1}
	A common alternative to the proposed summary statistics are what we call here {\em canonical summaries}, i.e., the 
	mean, variance, skewness, kurtosis and auto-correlation up to a certain lag (here 5) of the original data and their differences, the latter to account for temporal dependence. This leads to a 18-dimensional summary vector which we denote by $s_1(\tilde y_\theta)$ for a dataset $\tilde y_\theta$, and will only be considered in Section \ref{sec:impact_ingredients_algorithm} and Appendix \ref{AppExtraA}. 
\end{remark}

\subsubsection{Choice of the distance measure}

As a distance measure between the summaries of the observed dataset $y$ and synthetic dataset $\tilde y_\theta$ (cf. Lines 7 and 22 of Algorithm~\ref{alg:SMC_SBP_ABC}), we consider
\begin{equation}\label{eq:dist}
	d(s(y),s(\tilde{y}_\theta)):=\text{IAE}(\hat{S}_y,\hat{S}_{\tilde{y}_\theta}) + \alpha \text{IAE}(\hat{f}_y,\hat{f}_{\tilde{y}_\theta}),
\end{equation}
where $\text{IAE}$ is the \textit{integrated absolute error} between two $\mathbb{R}$-valued functions $g_1$, $g_2$ given by
\begin{equation}\label{eq:IAE}
	\text{IAE}(g_1,g_2):=\int_{\mathbb{R}} | g_1(x)-g_2(x) | \ dx  \geq 0.
\end{equation}
The constant $\alpha \geq 0$ in \eqref{eq:dist} corresponds to a \textit{weight} guaranteeing that the two IAEs have a \lq\lq comparable influence\rq\rq \ on the distance $d$ \cite{Buckwar2019}, which is relevant, as the invariant spectral density does not integrate to one. Here, the weight $\alpha$ is chosen as the magnitude of the area below the estimated spectral density $\hat{S}_y$ of the observed dataset $y$. Moreover, since $\hat{S}_y$, $\hat{S}_{\tilde{y}_\theta}$, $\hat{f}_y$ and $\hat{f}_{\tilde{y}_\theta}$ are estimated at discrete data points, the $\text{IAE}$ \eqref{eq:IAE} is approximated using rectangular integration. 

\begin{remark}
	The Wasserstein distance between $y$ and $\tilde y_\theta$, as proposed in \cite{ABCWass}, may be used instead of $\textrm{IAE}(\hat f_y,\hat f_{\tilde y_\theta})$, leading to marginal differences in the ABC posterior results (figures not shown). Moreover, as distance associated with the {\em canonical summary statistics} $s_1(\cdot)$, we consider the weighted Euclidean distance $d_1(s_1(y),s_1(\tilde y_\theta))=||\left(s_1(y)-s_1(\tilde y_\theta)\right)/\alpha_1||$, where $\alpha_1$ denotes a vector of weights with $\min(\alpha_1)>0$, the division by $\alpha_1$ being interpreted component-wise \cite{Prangle2017}. The vector $\alpha_1$ is computed as mean absolute deviations of all summaries obtained during the first iteration of Algorithm \ref{alg:SMC_SBP_ABC} (i.e., the acceptance-rejection step), as done, e.g., in \cite{PicchiniTamborrino2022}.  Other possible weights can be chosen (e.g. median  absolute deviation \cite{Csilleryetal2012}), all motivated by the idea of  normalising the effect caused by having summary statistics varying on very different scales \cite{Prangle2017}.
\end{remark}

\vspace{0.2cm}
\subsubsection{Choice of prior distribution}\label{sectionprior}

The prior distribution is relevant in the acceptance-rejection step of Algorithm \ref{alg:SMC_SBP_ABC} ($r=1$, see Line 4), and later on, when updating the weights at the $r$th iteration (cf. Line 25). Here, we use two sets of uniform priors
\begin{eqnarray}
	\label{eq:initial_prior_FHN}
	&&	\epsilon \sim U(0.01,0.5), \quad \gamma|\epsilon \sim U \bigl(\frac{\epsilon}{4},6 \bigr) 
	\quad \beta \sim U(0.01,6), \quad \sigma \sim U(0.01,1),
	\\
	\label{eq:initial_prior_FHN_ratdata}
	&&	\epsilon \sim U(0.01,1), \quad \gamma|\epsilon \sim U \bigl(\frac{\epsilon}{4},10 \bigr),
	\quad \beta \sim U(0.01,10), \quad \sigma \sim U(0.01,3),
\end{eqnarray}
with the former used for the simulation study (in Section \ref{sec:4_FHN}) and the latter for the real data study (in Section \ref{sec:5_FHN}). Sampling $\gamma$ from a uniform in $(\epsilon/4, \cdot)$ guarantees that the condition $\kappa>0$ is met, see Remark \ref{remark2}.

The use of continuous uniform priors is inspired by \cite{Buckwar2019}. However, other choices are possible. In Appendix \ref{AppendixPrior}, we also consider log-normal and exponential priors, and illustrate that their impact on the estimation results is marginal.

\vspace{0.2cm}
\subsubsection{Proposal Sampler}\label{SecpropSampler}

At the $r$th iteration, $\theta_j$, sampled from the weighted set $\{(\theta_{r-1}^{(j)},\omega_{r-1}^{(j)})_{j=1}^N\}$, is perturbed via a proposal sampler $K_r(\cdot|\theta_j)$ to obtain $\theta_j^*$ (cf. Lines 17 and 18 of Algorithm \ref{alg:SMC_SBP_ABC}).
Following \cite{Filippi2013}, we choose $K_r(\cdot|\theta_j)$ to be 
a multivariate normal perturbation kernel
with mean $\theta_j$ and covariance matrix given by twice the empirical weighted covariance matrix $\hat \Sigma_{r-1}$ of the population $\{(\theta_{r-1}^{(j)},\omega_{r-1}^{(j)})_{j=1}^N\}$, i.e.  $K_r(\cdot|\theta_j)=\mathcal{N}(\theta_j,2\hat\Sigma_{r-1})$. Hence, $k_r(\cdot|\theta_{r-1}^{(l)})$ in 
Line 25 of Algorithm \ref{alg:SMC_SBP_ABC} is then the density of 
$\mathcal{N}(\theta_{r-1}^{(l)},2\hat\Sigma_{r-1})$    
evaluated in $\theta_r^{(j)},$ for $j,l\in\{1,\ldots,N\}$. 

This Gaussian proposal sampler, which we call here \texttt{standard}, is obviously not the only possible choice. In Appendix \ref{AppendixA}, we also consider the {\em optimal local covariance matrix} (\texttt{olcm}) Gaussian proposal sampler introduced in \cite{Filippi2013}, and compare the impact of the two samplers on the estimation results. 

Differently than from the prior, the condition $\kappa>0$ is not directly accounted for in the proposal samplers $K_r$. However, perturbed samples $\theta^{*}$ which do not satisfy $\kappa>0$ or lie outside the support of the prior (i.e. $\pi(\theta^*)=0$) are immediately rejected and replaced by new ones (cf. Line 19 of Algorithm \ref{alg:SMC_SBP_ABC} and the right panel of Figure \ref{runtime10m} in Appendix \ref{AppExtraA}). This is a standard procedure in sequential ABC samplers, see e.g. Step 4 of the SMC-ABC Algorithm~1 in \cite{Sun2015} as well as Line 18 of the Sequential Important Sampling ABC (SIS-ABC) Algorithm~1 and Line 19 of the SMC-ABC Algorithm~2 in \cite{PicchiniTamborrino2022}. Immediate rejections due to entries of $\theta^{*}$ exceeding the upper bound of the respective uniform prior's support can be avoided by using priors with support $(0,\infty)$ instead (cf. Appendix \ref{AppendixPrior}). Rejections due to entries of $\theta^{*}$ being negative may be avoided by adjusting the algorithm to account for logged parameters. Both adjustments have a marginal impact on the estimation results and computing times. Also the number of particles sampled from $K_r$ with $\kappa\leq 0$ and thus immediately disregarded is negligible.

\subsubsection{Threshold levels}\label{secthreshold}
The first tolerance level $\delta_1$, needed in the first iteration of the SMC-ABC Algorithm \ref{alg:SMC_SBP_ABC} (the acceptance-rejection step), is obtained via an ABC pilot study as the $p$th percentile of $10^4$ distances between the summaries of the reference data and of the datasets generated with  
parameter values sampled from the prior. All following tolerance levels $\delta_r$, $r>1$, are automatically chosen
as the $p$th percentile of the $N$ accepted distances computed at the previous iteration. Alternatively, one may choose the new tolerance levels as a percentile of all previous distances (both rejected and accepted), as done in \cite{PicchiniTamborrino2022}, or via a procedure which, progressively, albeit slowly, reduces them while maintaining a reasonably high effective sample size (ESS), as suggested in \cite{DelMoral2012}.


\subsection{Implementation details}
\label{sec:3:impl.det.}

All algorithms are implemented in the software environment \texttt{R} \cite{R}, using the package \texttt{Rcpp} \cite{Rcpp}, offering a seamless integration of \texttt{R} and \texttt{C++}. In particular, Algorithm \ref{alg:Splitting} is coded in \texttt{C++}, which speeds up the 
synthetic data generation significantly, reducing thus the computational cost of the ABC procedure. 
Taking advantage of the independence between the $N$ particles in the \textit{for loops} in Algorithm \ref{alg:SMC_SBP_ABC}, the code is parallelised using the packages \texttt{foreach} and 
\texttt{doParallel}. The R-package \texttt{mvnfast} \cite{Fasiolo} is used to draw samples from the multivariate normal perturbation kernel (using the function \texttt{rmvn}) and to compute its density (using the function \texttt{dmvn}).

The invariant spectral density $S_{\tilde y_\theta}$ and invariant density $f_{\tilde y_\theta}$ for a dataset $\tilde y_\theta$, used as summaries in \eqref{s}, are estimated via the R-functions \texttt{spectrum} and \texttt{density}, respectively. Within the function \texttt{spectrum}, we avoid a logarithmic scale by setting \texttt{log}=``no''. Within the function \texttt{density}, we use the default value for the bandwidth \texttt{bw} and set the number of points at which the density is estimated to \texttt{n}$=10^3$. 

The number of particles accepted in each iteration is fixed to $N=10^3$ in all experiments. The threshold levels are chosen as the $50$th percentile (median) of the simulated (for the pilot study) or accepted (for $r \geq 1$) distances. All synthetic datasets (cf. Lines 5 and 20 of Algorithm \ref{alg:SMC_SBP_ABC}) are generated with Algorithm~\ref{alg:Splitting}, using 
the step-size $\Delta=0.02$. 

An easy-to-grasp implementation of the SBP SMC-ABC method (i.e., Algorithm~\ref{alg:SMC_SBP_ABC} with the ingredients presented in Section~\ref{sec:3:ABCingredients}, including Algorithm \ref{alg:Splitting} for synthetic data simulation) is  available at \href{https://github.com/IreneTubikanec/SMC-ABC_FHN}{https://github.com/IreneTubikanec/SMC-ABC\_FHN}. Moreover, a comprehensive \texttt{R}-package, providing the codes for running Algorithm~\ref{alg:SMC_SBP_ABC} with different summary statistics (structure-based and canonical), priors (uniform, log-normal and exponential),  distances (IAE and Wasserstein) and proposal samplers (Gaussian samplers \texttt{standard} and \texttt{olcm}, described in Section~\ref{SecpropSampler} and Appendix~\ref{AppendixA}, respectively) is  available at \href{https://github.com/massimilianotamborrino/SMCABCFHN}{https://github.com/massimilianotamborrino/SMCABCFHN}.

All case studies, based on both simulated and real data, have been run on multiple core High Performance Computing (HPC) clusters   
at both the University of Klagenfurt and the University of Warwick. Unless differently stated, all reported runtimes were obtained by launching the relevant codes on $48$ parallel cores of a HPC cluster at the University of Warwick (Dell PowerEdge C6420, 2.9 GHz, 24-core processors). 


\section{Inference of the FHN parameters from simulated data}
\label{sec:4_FHN}

In this section, we investigate the performance of the proposed statistical procedure for parameter estimation in the stochastic FHN model \eqref{FHN}. In~Section \ref{SectionEqui}, we analyse the SBP SMC-ABC method, based on different simulated observed datasets. In Section \ref{sec:impact_ingredients_algorithm}, we study the impact of summary statistics on the ABC results, illustrating how the structure-based summaries may outperform the canonical ones. In Appendix \ref{AppExtraA}, we investigate the computational cost of the different summaries. In Appendix \ref{AppendixPrior}, we illustrate the marginal impact of the prior distribution on the ABC results. Moreover, in Appendix \ref{AppendixA}, we also extensively investigate the impact of proposal samplers in terms of acceptance rate, ESS, quality of the inference and running times. Other ingredients in the procedure, such as the percentile in defining the threshold levels (in particular, 10, 25, 50\%) and alternative distances between the summaries (namely, the Wasserstein distance between the invariant densities instead of the IAE) have shown to have a limited impact on the results, and are therefore not reported here.

Throughout this section (and the appendix), the underlying true parameter value of $\theta$ \eqref{eq:theta} used to generate the respective observed dataset is the same as in \cite{Ditlevsen2019}, namely
\begin{equation*}\label{eq:theta_true_FHN}
	\theta=(\epsilon,\gamma,\beta,\sigma)=(0.1,1.5,0.8,0.3).
\end{equation*}


\subsection{Inference under different observation settings}\label{SectionEqui}

\begin{figure}[t]
	\begin{centering}
		\includegraphics[width=0.9\textwidth]{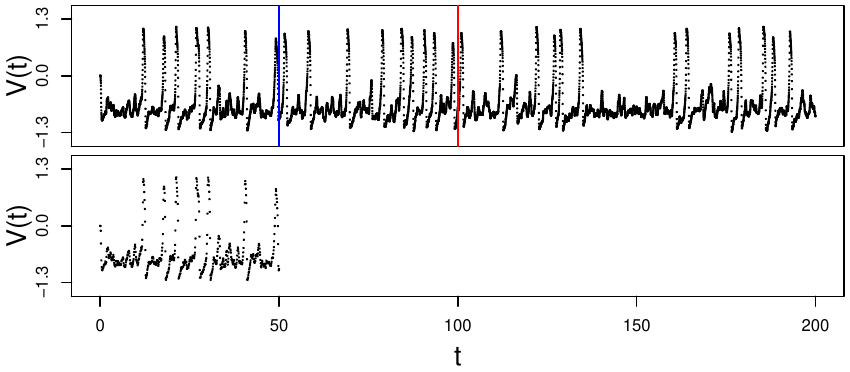}	
		\caption{Top panel: Observed (simulated) dataset for $\Delta_\textrm{obs}=0.02$ and $T_\textrm{obs}=50$ (yielding $n=2500$ data points, up to the blue line), $T_\textrm{obs}=100$ ($n=5000$, up to the red line) and $T_\textrm{obs}=200$ ($n=10^4)$. Bottom panel: Observed (simulated) dataset for $\Delta_\textrm{obs}=0.08$ and  $T_\textrm{obs}=50$ (yielding $n=625$ data points).}
		\label{fig:ABC_oberseved_data_equi}
	\end{centering}
\end{figure}

The performance of the proposed SBP SMC-ABC method is investigated under different observation regimes, depending on the choice of the observation time horizon $T_\textrm{obs}$, and the observation time step~$\Delta_\textrm{obs}$. In particular, we consider $T_\textrm{obs}=50, 100, 200$ and $\Delta_\textrm{obs}=0.02, 0.04, 0.08$, yielding 
observed datasets with a size ranging from $n=10^4$ (for $T_\textrm{obs}=200$ and $\Delta_\textrm{obs}=0.02$) to only $n=625$ (for $T_\textrm{obs}=50$ and $\Delta_\textrm{obs}=0.08$) data points.

All observed datasets are obtained via Algorithm \ref{alg:Splitting}, using a simulation step-size $\Delta=10^{-4}$ and time horizon $T=200$. They are then sub-sampled with the corresponding observation time step $\Delta_\textrm{obs}$ and cut to the respective observation length $T_\textrm{obs}$, as visualised in Figure \ref{fig:ABC_oberseved_data_equi}. The synthetic datasets are instead obtained with $T=T_\textrm{obs}$ and $\Delta=0.02$, and then sub-sampled with the corresponding $\Delta_\textrm{obs}$. In this section, the computational budget of Algorithm \ref{alg:SMC_SBP_ABC} is set to $\textrm{Nsim}_\textrm{max}=10^6$ synthetic data simulations. Moreover, the results reported here are based on one trial per observation setting. However, we verified that they are consistent across repeated runs (figures not shown).

\begin{figure}[t]
	\begin{centering}
		\includegraphics[width=1.0\textwidth]
		{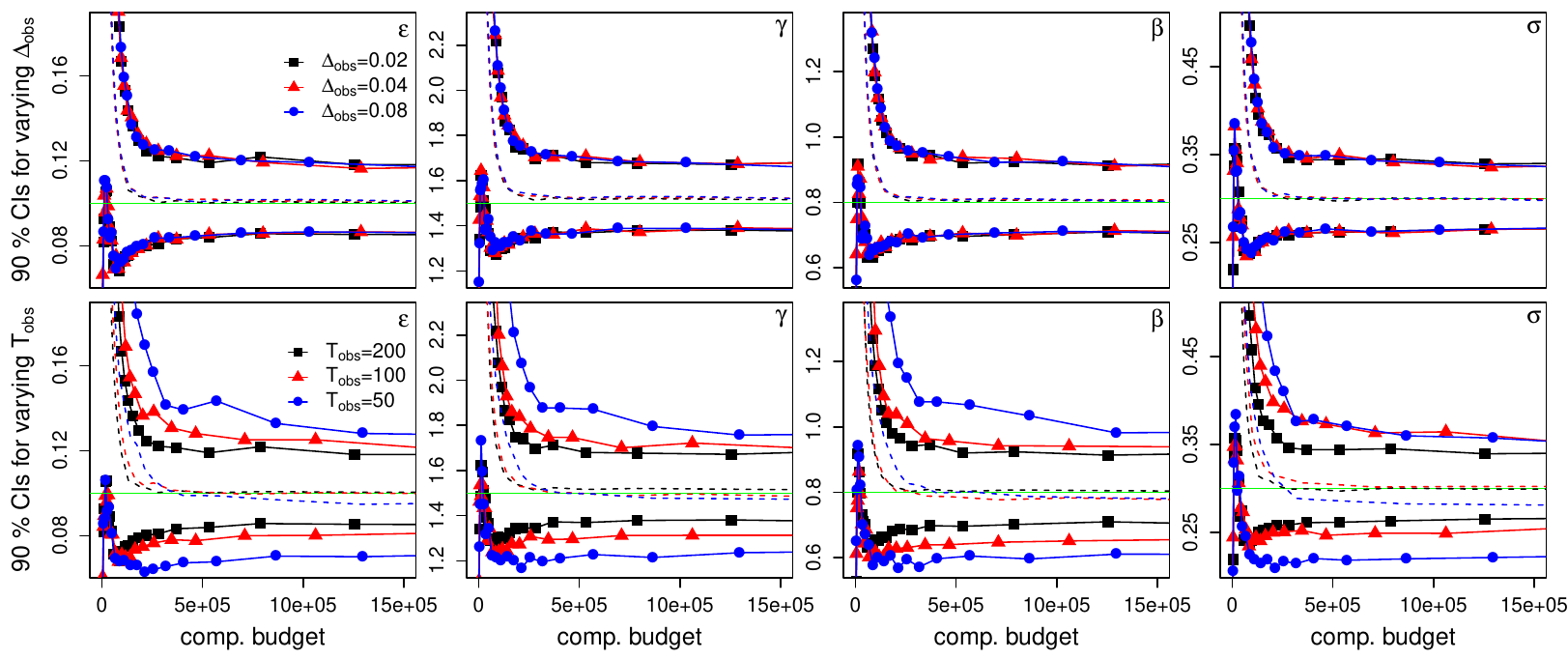}
		\caption{$90$\% CIs of the marginal posterior densities of the parameter vector $\theta$ \eqref{eq:theta} of the stochastic FHN model \eqref{FHN} as a function of the computational budget. Top panels: $T_\textrm{obs}=200$ and $\Delta_\textrm{obs}=0.02$ (black squares), $0.04$ (red triangles), $0.08$ (blue circles). Bottom panels: $\Delta_\textrm{obs}=0.02$ and $T_\textrm{obs}=200$ (black squares), $100$ (red triangles), $50$ (blue circles). The green horizontal lines and the dashed lines indicate the true parameter values and the 
			corresponding posterior medians,~respectively. The number of observed data points is the same (for the same colours and shapes) for the top and bottom panels.}
		\label{fig:ABC_CIs}
	\end{centering}
\end{figure}

\begin{table}[h!]
	{\small  		\caption{ABC posterior means and standard deviations of $\theta$ \eqref{eq:theta}, under different observation settings.  
			The true parameter values are $\epsilon=0.1$, $\gamma=1.5$, $\beta=0.8$ and $\sigma=0.3$. The reported runtimes are medians across ten runs of the algorithm.}
		\vspace{-0.5cm}
		\label{table:ABCresults}
		\begin{center}
			\scalebox{0.95}{
				\begin{tabular}{cccccc}
					\hline 
					$T_\textrm{obs}$ & $\Delta_\textrm{obs}$ & $n$ & ABC posterior means & ABC posterior stand. dev.s & Median runtime
					\\ \hline \hline 
					$200$ & $0.02$ & $10^4$ & $(0.101,1.523,0.808,0.300)$ & $(0.010,0.087,0.062,0.023)$ &3 min 7 s\\
					
					$200$ & $0.04$ & $5000$ & $(0.101,1.524,0.808,0.300)$ & $(0.009,0.086,0.061,0.022)$ &2 min 3 s\\
					
					$200$ & $0.08$ & $2500$ & $(0.102,1.531,0.812,0.301)$ & $(0.010,0.089,0.064,0.023)$ &1 min  41 s\\ 
					
					\hline 
					
					$100$ & $0.02$ & $5000$ & $(0.101,1.504,0.787,0.304)$ & $(0.014,0.119,0.085,0.034)$&2 min 25 s \\ 
					
					$100$ & $0.04$ & $2500$ & $(0.102,1.508,0.789,0.305)$ & $(0.014,0.125,0.089,0.033)$ &1 min 47s\\
					
					$100$ & $0.08$ & $1250$ & $(0.101,1.506,0.787,0.304)$ & 
					$(0.013,0.116,0.084,0.033)$ &1 min 26 s\\
					
					\hline 
					
					$50$ & $0.02$ & $2500$ & $(0.096,1.480,0.787,0.284)$ & $(0.017,0.160,0.113,0.040)$&1 min 25 s \\ 
					
					$50$ & $0.04$ & $1250$ & $(0.098,1.486,0.789,0.281)$ & $(0.018,0.166,0.125,0.039)$ &1 min 8 s\\
					
					$50$ & $0.08$ & $625$ & $(0.098,1.490,0.793,0.282)$ & $(0.018,0.171,0.123,0.041)$ &1 min 3 s\\ 
					\hline
			\end{tabular}}
	\end{center}}
\end{table}

\paragraph{ABC results}

\begin{figure}[t]
	\begin{centering}
		\includegraphics[width=1.0\textwidth]
		{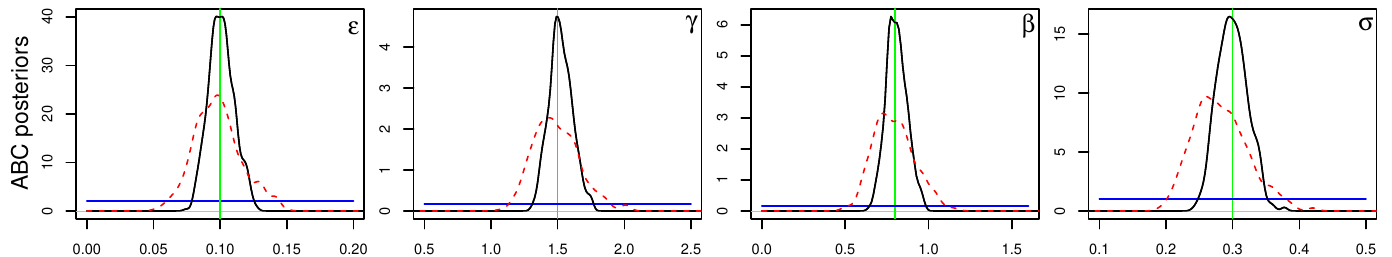}
		\caption{Marginal posterior densities of the parameter vector $\theta$~\eqref{eq:theta} of the stochastic FHN model~\eqref{FHN} (visualised over only a part of the prior domain). 
			Black solid: $T_\textrm{obs}=200$, $\Delta_\textrm{obs}=0.02$, $n=10^4$. Red~dashed: $T_\textrm{obs}=50$, $\Delta_\textrm{obs}=0.08$, $n=625$.  The green vertical lines and the blue horizontal lines indicate the true parameter values and the prior densities \eqref{eq:initial_prior_FHN}, respectively.}
		\label{fig:ABC_posteriors}
	\end{centering}
\end{figure}

\begin{figure}[h!]
	\begin{centering}
		\includegraphics[width=1.0\textwidth]
		{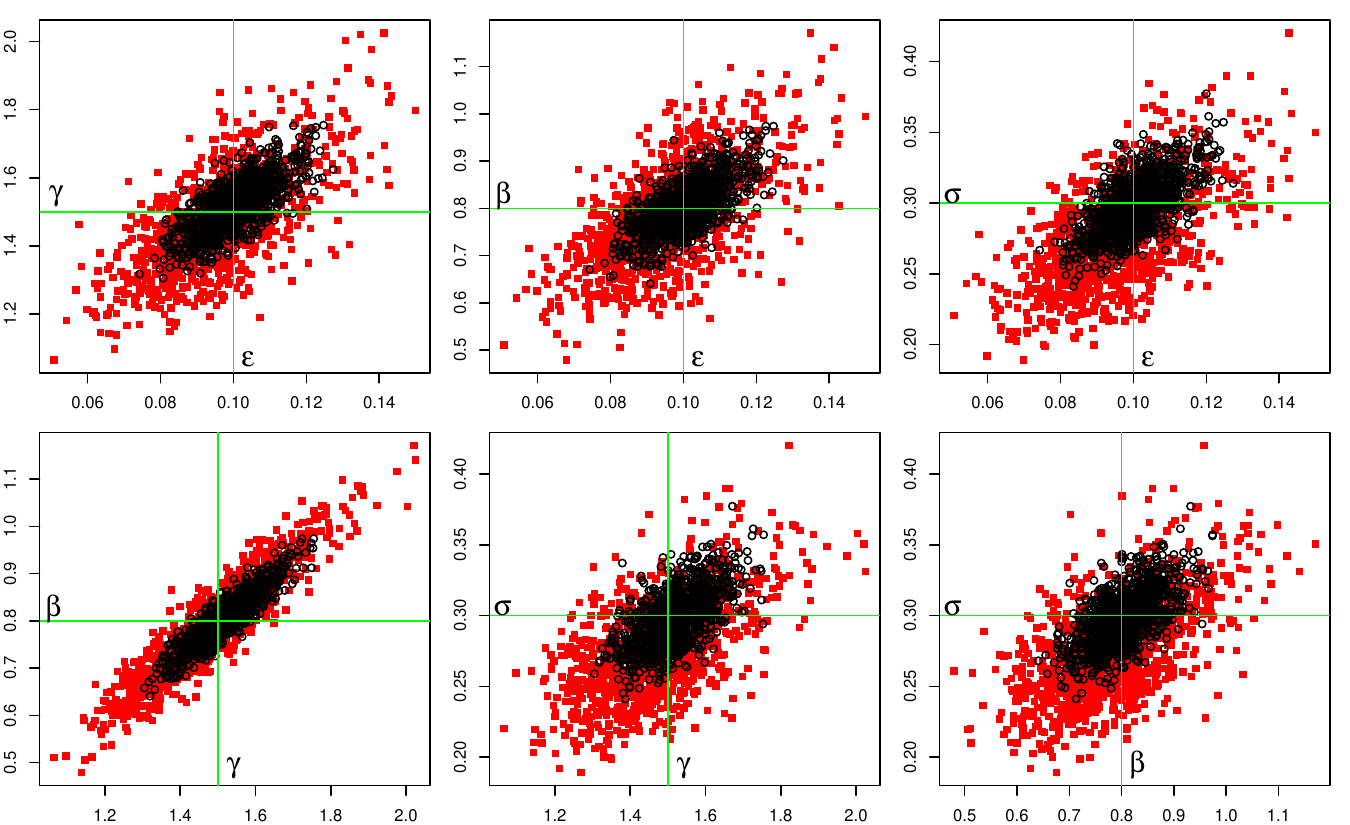}
		\caption{Scatter plots of pairs of the kept ABC posterior samples.
			Black dots: $T_\textrm{obs}=200$, $\Delta_\textrm{obs}=0.02$, $n=10^4$. Red~dots: $T_\textrm{obs}=50$, $\Delta_\textrm{obs}=0.08$, $n=625$.  The green lines  indicate the respective true parameter values.}
		\label{fig:ABC_posteriors_scatter}
	\end{centering}
\end{figure}

Our inferential results show that the observation time step $\Delta_\textrm{obs}$ does not significantly influence the performance of the SBP SMC-ABC method,  when choosing a (large enough) observation time horizon $T_\textrm{obs}$. This becomes evident when looking at the top panels of Figure~\ref{fig:ABC_CIs}, which show the $90$\% credible intervals (CIs) of the derived marginal ABC posterior densities of $\theta$~\eqref{eq:theta} as a function of the computational budget, when fixing $T_\textrm{obs}=200$ and considering $\Delta_\textrm{obs}=0.02$ (black squares), $\Delta_\textrm{obs}=0.04$ (red triangles) and $\Delta_\textrm{obs}=0.08$ (blue circles). 
This is also the case when choosing $T_\textrm{obs}=50$ or $T_\textrm{obs}=100$.

In contrast to that, the observation time horizon $T_\textrm{obs}$ does influence the inferential results when fixing the time-step $\Delta_\textrm{obs}$, with smaller $T_\textrm{obs}$ yielding broader CIs. This can be observed in the bottom panels of Figure \ref{fig:ABC_CIs}, where we fix $\Delta_\textrm{obs}=0.02$ and consider 
$T_\textrm{obs}=200$ (black squares), $T_\textrm{obs}=100$ (red triangles) and $T_\textrm{obs}=50$ (blue circles). Similar considerations hold when choosing $\Delta_\textrm{obs}=0.04$, or $\Delta_\textrm{obs}=0.08$. 

It is important to stress that the number of observed data points $n$ is the same (for the same colours and shapes) in the top and bottom panels of Figure \ref{fig:ABC_CIs}, i.e., $n=2500$ (blue circles), $n=5000$ (red triangles) and $n=10000$ (black squares). However, the increased sample size leads to a more accurate ABC inference (in the sense of narrower CIs) only when increasing the time horizon. This is also confirmed by Table \ref{table:ABCresults}, where the ABC posterior means and standard deviations obtained under the different observation settings are reported together with the median runtimes obtained over ten runs of the SBP SMC-ABC algorithm. These findings are in agreement with the requirement $T_\textrm{obs}\to \infty$ for contrast estimators of SDEs to be consistent and asymptotically normal (see, e.g., \cite{Ditlevsen2019}).

Overall, all computed ABC posterior densities are unimodal, cover the true parameter values (green horizontal lines in Figure \ref{fig:ABC_CIs}) with  posterior medians (dashed lines in Figure \ref{fig:ABC_CIs}) close to them (see also Table \ref{table:ABCresults}, where the respective posterior means are reported). Only for $T_\textrm{obs}=50$, we observe that the ABC posterior medians (resp. the means) of the noise parameter $\sigma$ are slightly off the true value of $0.3$.

In Figure \ref{fig:ABC_posteriors}, we report the marginal ABC posterior densities of $\theta$ \eqref{eq:theta}, derived from the two most extreme observation settings: (i) $T_\textrm{obs}=200$, $\Delta_\textrm{obs}=0.02$, $n=10^4$ (black solid lines); (ii) $T_\textrm{obs}=50$, $\Delta_\textrm{obs}=0.08$, $n=625$ (red dashed lines). In both scenarios, the marginal posterior densities show a clear update compared to the respective uniform prior densities (blue horizontal lines) and cover the true parameter values (green vertical lines), with the posteriors of scenario (i) being narrower than those of scenario (ii) (cf. the values for the corresponding ABC posterior standard deviations in Table \ref{table:ABCresults}).

To provide a more comprehensive picture of the obtained ABC posterior distributions, in Figure~\ref{fig:ABC_posteriors_scatter}, we also report scatter plots of pairs of the kept ABC posterior samples for scenarios (i) (black dots) and (ii) (red dots), with similar results obtained in the other settings. The underlying true parameter values, which are represented by the green vertical and horizontal lines, are covered by the respective point clouds. All pairs of parameters show some correlation, with $\gamma$ and $\beta$ strongly correlated, having a correlation coefficient of $0.912$ and $0.926$ for scenario (i) and (ii), respectively.

\vspace{0.3cm}
\subsection{Inference under different summary statistics}
\label{sec:impact_ingredients_algorithm}
\vspace{0.1cm}

While the superiority and necessity of using structure-preserving numerical schemes over standard (non-preserving) simulation methods (such as Euler-Maruyama) within simulation-based algorithms have been intensively investigated in \cite{Buckwar2019}, no comparison between structure-based summaries~\eqref{s} and \lq\lq canonical\rq\rq\ summaries (cf. Remark \ref{rem:s1}) has been provided yet. Here, we~close~this~gap. 

\begin{figure}[t]
	\begin{centering}
		\includegraphics[width=1.0\textwidth]{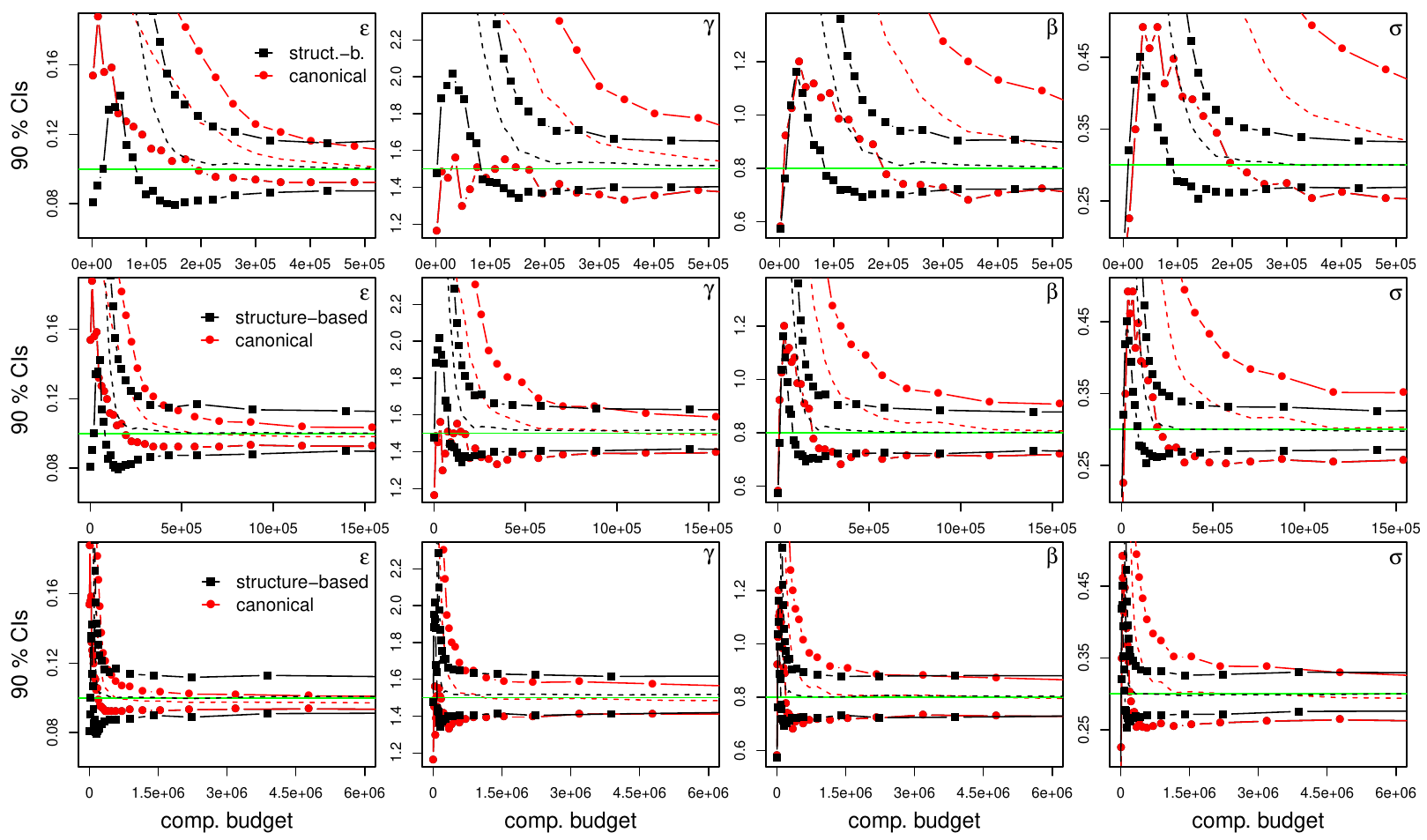}
		\caption{90\% CIs of the marginal posterior densities of the parameter vector $\theta$ \eqref{eq:theta} of the stochastic FHN model \eqref{FHN} obtained from Algorithm \ref{alg:SMC_SBP_ABC} with either structure-based (cf. \eqref{s}, black lines) or canonical 
			(cf. Remark~\ref{rem:s1}, red lines) summaries as a function of the computational budget. The green horizontal lines and the dashed lines indicate the true parameter values and the corresponding posterior medians, respectively.
		} 
		\label{FigABC_summaries}
	\end{centering}
\end{figure}

In Figure \ref{FigABC_summaries}, we report the 90\% CIs of the marginal posterior densities obtained 
by applying Algorithm~\ref{alg:SMC_SBP_ABC} to the observed data visualized in Figure \ref{fig:ABC_oberseved_data_equi} (observation setting $T_\textrm{obs}=200$, $\Delta_\textrm{obs}=0.02$), using the canonical (red lines) or the structure-based  (black lines) summaries (the latter corresponding to the SBP SMC-ABC method). 
These CIs are plotted as a function of the computational budget, ranging up to $5\cdot 10^5$ (top panels), $1.5\cdot 10^6$ (middle panels) or  $6\cdot 10^6$ (bottom panels). 
For $\beta$ and $\sigma$, the CIs obtained when using the canonical summaries are wider than those derived from the structure-based summaries up to circa $2$-$3$ million model simulations, when they
have approximately the same width. Moreover, using structure-based summaries yields CIs approaching the desired posterior region much faster, centered around the true parameter values for as few as $200$-$300$ thousand simulations, and remaining almost unchanged when increasing the computational budget. This is not the case when using the canonical summaries. On the one hand, they lead to posteriors with CIs for $\epsilon$ and $\sigma$ continuing shrinking when increasing the number of simulations (see the bottom panels). On the other hand, they yield posterior medians much larger than the true parameter values under small computational budgets (see the top panels), and below the true values under larger computational budgets (see in particular the bottom left panel for $\epsilon$).  

\begin{figure}[h!]
	\includegraphics[width=1.0\textwidth]{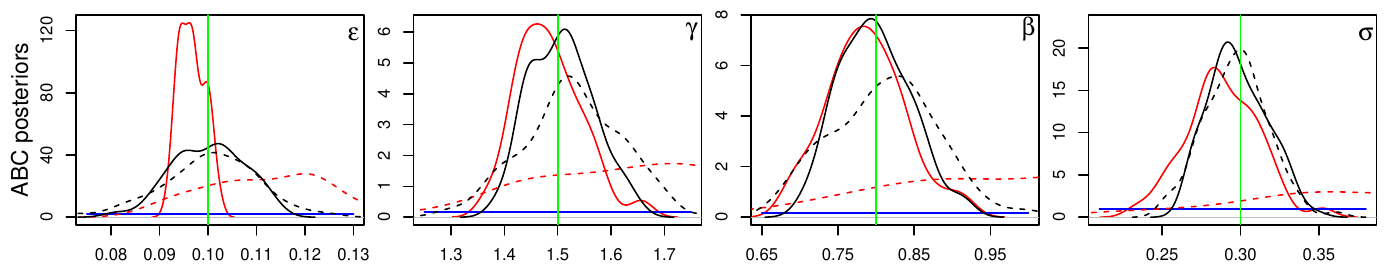}
	\caption{Marginal posterior densities of the parameter vector $\theta$ \eqref{eq:theta} of the stochastic FHN model~\eqref{FHN} obtained from the SMC-ABC Algorithm \ref{alg:SMC_SBP_ABC} with either structure-based (cf. \eqref{s}, black lines) or canonical (cf. Remark~\ref{rem:s1}, red lines)   summaries when using approximately $3\cdot 10^5$  (dashed lines) and $11$-$13$ million (solid lines) model simulations, respectively. The green vertical lines and the blue horizontal
		lines indicate the true parameter values and the prior densities, respectively.}
	\label{FigABCposteriors_summaries}
\end{figure}

Similar observations can be made from Figure \ref{FigABCposteriors_summaries}, where we plot the  
marginal posterior densities for a computational budget of approximately $3 \cdot 10^5$ (dashed lines) and 12 million (solid lines) model simulations. 
While the structure-based summaries (black lines) yield satisfactory results even under a small computational budget, the canonical summaries (red lines) fail to do so. 
The results for $\epsilon$ are particularly problematic, suggesting that the canonical  summaries may be driving the posterior mass to a narrow and \lq\lq wrong\rq\rq \ area when increasing the computational budget, as also observed from the bottom left panel in Figure \ref{FigABC_summaries} and the ABC posterior means and standard deviations reported in Table \ref{table:ABCresults_summaries}.

\begin{table}[t]
	{\small  
		\caption{ABC posterior means and standard deviations of $\theta$ \eqref{eq:theta}, obtained from the SMC-ABC Algorithm \ref{alg:SMC_SBP_ABC} with either structure-based (cf. \eqref{s}) or canonical (cf. Remark~\ref{rem:s1})  summaries when using approximately $11-13$ million model simulations.}
		\vspace{-0.5cm}
		\label{table:ABCresults_summaries}
		\begin{center}
			\scalebox{0.95}{
				\hspace{-1.0cm}
				\begin{tabular}{ccc}
					\hline 
					Summaries & ABC posterior means & ABC posterior stand. dev.s  \\ \hline 
					Structure-based & $( 0.100, 1.509, 0.797, 0.297)$ & $( 0.007, 0.069, 0.050, 0.017)$  \\
					Canonical  & $(0.097, 1.482, 0.792, 0.292)$ & $(0.002, 0.051, 0.042, 0.019)$ \\
					\hline
			\end{tabular}}
	\end{center}}
\end{table}

In Appendix \ref{AppExtraA}, we also investigate the computational cost of the summaries, both in terms of evaluation time per dataset (in the order of milliseconds) and runtimes of the corresponding algorithms as functions of the computational budget. Calculating canonical summaries for a given dataset is slightly faster than computing the structure-based ones, but the overall impact on the runtime is negligible. Considering that the algorithm based on structure-based summaries requires fewer model simulations to achieve a desired accuracy, these summaries are then preferable to the canonical ones also in terms of overall algorithm runtimes.  

\begin{remark}
	Interestingly, considering the alternative structure-based summaries $s(\tilde y_\theta)=(\hat{f}_{\tilde y_\theta},\hat{R}_{\tilde y_\theta})$ obtained by replacing the spectral density with the underlying auto-correlation function (here estimated for time-lags $1,\ldots,25$) does not yield inferential results as good as those obtained with the proposed structure-based summaries \eqref{s}.
	While the resulting ABC posterior distributions also cover the true parameter values well, their standard deviations are slightly larger and their means slightly more off the true values compared to the results obtained via \eqref{s} (figures not shown). Considering $s(\tilde y_\theta)=(\hat{f}_{\tilde y_\theta},\hat{S}_{\tilde y_\theta},\hat{R}_{\tilde y_\theta})$ (i.e., adding the auto-correlation function to \eqref{s}) does not yield a notable improvement. This may be explained by the fact that the spectral density already uses the information contained in the auto-correlation function.
\end{remark}


\section{Inference of the FHN parameters from real action potential data}
\label{sec:5_FHN}

In this section, we infer the parameters $\theta$ \eqref{eq:theta} of the stochastic FHN model \eqref{FHN} from real data consisting of membrane voltages recorded in rats.


\subsection{Description of the data}

The investigated data have been obtained via a neural recording from the 5th lumbar dorsal rootlet of an adult female Sprague Dawley rat, and are available at \url{https://data.mendeley.com/datasets/ybhwtngzmm/1}. A detailed description of the data and the corresponding experimental design are provided in \cite{METCALFE2020}. In brief, these data consist of twenty recordings of length $T_\textrm{obs}=250$ ms with a sampling rate of $50$ kHz (and thus $\Delta_\textrm{obs}=0.02$), ten obtained with the animal at rest (see the datasets in the folder ``Resting'') and ten during stimulation (of the dermatome, activating the corresponding axons in the dorsal root, see the datasets in the folder ``Cutaneous Stimulation''). Each recording consists of $5$ channels of voltage~data.

Here, we investigate four of the available twenty recordings, in particular the recordings \textit{1553}, \textit{1608} and \textit{1554}, \textit{1609}, stored in the folders ``Resting'' and ``Cutaneous Stimulation'', respectively, the filenames corresponding to the times the recordings were started. In each recording, we focus on Channel $1$, which represents the most distal electrode (i.e., the one closest to the tail). Similar results (not reported) are obtained when considering other channels and recordings. 


\subsection{Parameter inference}

\paragraph{ABC setup}

The stochastic FHN model \eqref{FHN} is fitted to these four datasets using the SBP SMC-ABC method, with a computational budget of $\textrm{Nsim}_\textrm{max}=2\cdot 10^6$.  All synthetic datasets are generated with step-size $\Delta=\Delta_{\textrm{obs}}=0.02$ and time horizon $T=T_{\textrm{obs}}=250$. 
Note that, the real datasets are centered close to zero, with a mean value of approximately $0.003$ (the empirical means of the datasets \textit{1553}, \textit{1608}, \textit{1554} and \textit{1609} are $0.00277$,  $0.00265$, $0.00271$ and $0.00263$, respectively). Therefore, in the following, all investigated datasets (real and synthetic) are centered.

\paragraph{Estimation results}

\begin{figure}[H]
	\begin{centering}
		\includegraphics[width=0.95\textwidth]
		{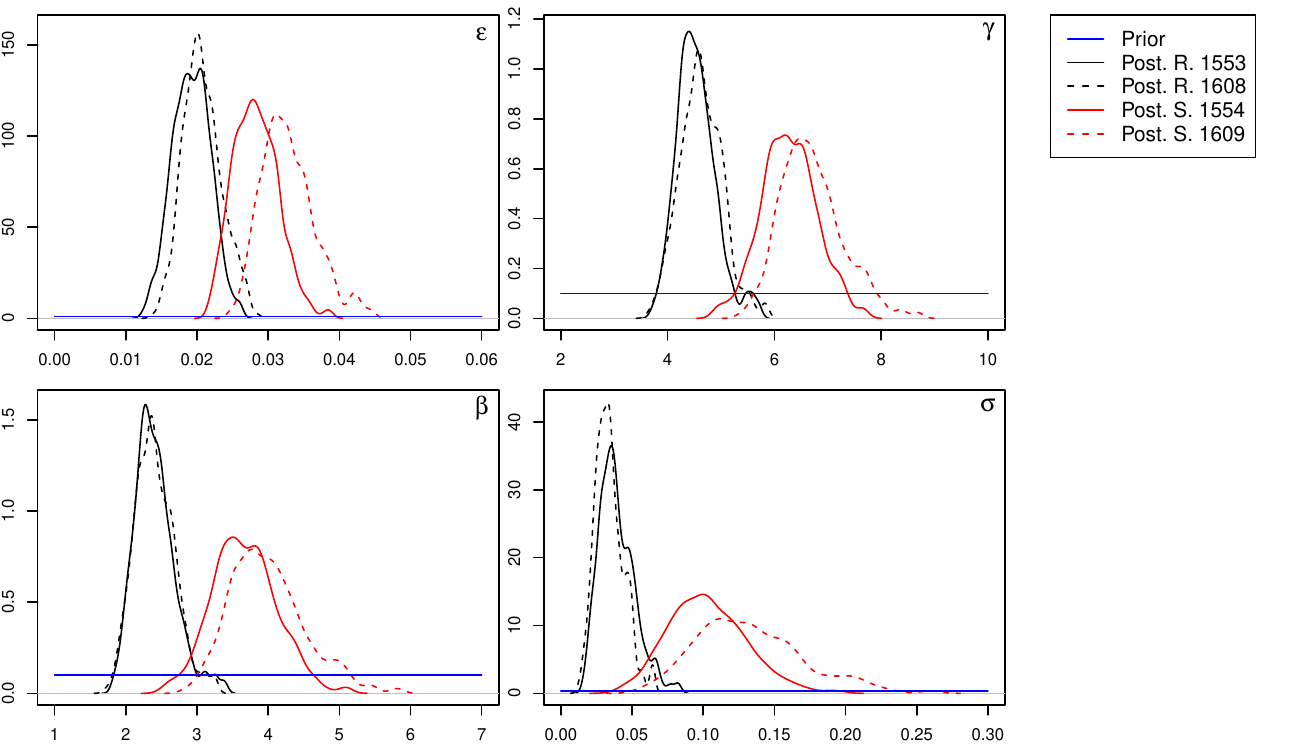}
		\caption{Marginal posterior densities of the parameter vector $\theta$ \eqref{eq:theta} of the stochastic FHN model~\eqref{FHN} fitted on the real action potential datasets \textit{1553} (animal at rest, black solid lines), \textit{1554} (animal stimulated, red solid lines), \textit{1608} (animal at rest, black dashed lines) and \textit{1609} (animal stimulated, red dashed lines), using the SBP SMC-ABC method. The horizontal blue lines indicate the uniform prior densities \eqref{eq:initial_prior_FHN_ratdata}.}
		\label{fig:ABCresults_ratdata}
	\end{centering}
\end{figure}

Figure \ref{fig:ABCresults_ratdata} shows the marginal ABC posterior densities obtained from the action potential recordings made when the animal was at rest (\textit{1553}: black solid lines, \textit{1608}: black dashed lines), and when it was stimulated (\textit{1554}: red solid lines, \textit{1609}: red dashed lines). The horizontal blue lines correspond to the underlying uniform prior densities, only a section of their domains being visualised. In addition, Table \ref{table:ABCresults_ratdata} reports relevant descriptive statistics of the obtained posterior distributions.     

All marginal posterior densities are unimodal, notably different from the priors. Interestingly, the posteriors for the two ``resting'' (resp. \lq\lq stimulating\rq\rq) scenarios are similar, while they differ 
across these two cases, with the ABC marginal posterior densities of all parameters shifted towards larger values 
in the ``stimulating'' scenarios compared to the ``resting'' scenarios (see also Table~\ref{table:ABCresults_ratdata}). Moreover, their uncertainty also increases under the ``stimulating'' settings, as confirmed by the larger  posterior standard deviations and the wider $90$\% CIs reported in Table~\ref{table:ABCresults_ratdata}. A deeper analysis of the derived posterior distributions is provided in Appendix \ref{AppendixScatter}.

\begin{table}[H]
	{\small  
		\caption{ABC posterior means, standard deviations and credible intervals  of $\theta$ \eqref{eq:theta}, obtained under four different real action potential datasets.}
		\vspace{-0.5cm}
		\label{table:ABCresults_ratdata}
		\begin{center}
			\scalebox{0.9}{
				\hspace{-1.0cm}
				\begin{tabular}{cccc}
					\hline 
					case & ABC posterior means & ABC post. stand. dev.s & $90\%$  credible intervals \\ \hline 
					Rest. \textit{1553} & $(0.019,4.535,2.407,0.039)$ & $(0.003,0.368,0.286,0.013)$ & $[0.015,0.023]$, $[3.991,5.197]$, $[2.007,2.934]$, $[0.022,0.065]$ \\
					Rest. \textit{1608} & $(0.021,4.629,2.406,0.034)$ & $(0.003, 0.397,0.281,0.010)$ & $[0.017,0.026]$, $[3.988,5.313]$, $[1.975,2.871]$, $[0.020,0.051]$ \\
					
					Stim. \textit{1554} & $(0.028,6.272,3.670,0.103)$ & $(0.003,0.521,0.454,0.028)$ & $[0.023,0.034]$, $[5.425,7.191]$, $[2.964,4.443]$, $[0.062,0.105]$ \\
					
					Stim. \textit{1609} & $(0.033,6.701,3.995,0.133)$ & $(0.004,0.589,0.529,0.038)$ & $[0.028,0.039]$, $[5.875,7.758]$, $[3.233,5.003]$, $[0.079,0.204]$ \\
					\hline
			\end{tabular}}
	\end{center}}
\end{table}

\paragraph{Fitted FHN model}

\begin{figure}[H]
	\begin{centering}
		\includegraphics[width=0.92\textwidth]
		{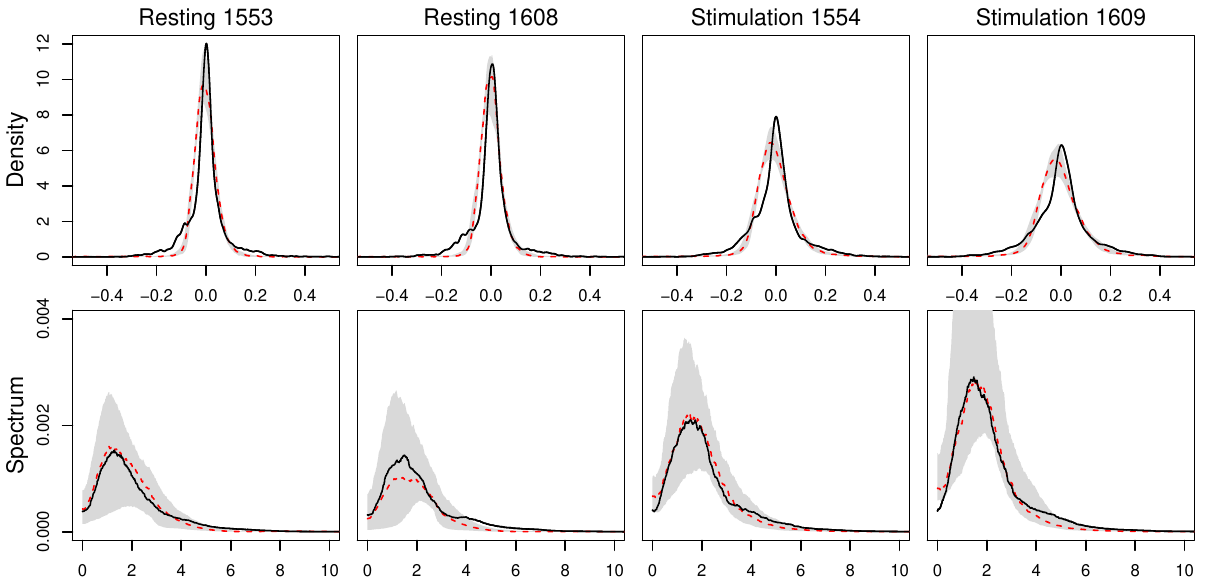}
		\caption{Structure-based summaries \eqref{s} of the action potential recordings \textit{1553}, \textit{1554}, \textit{1608}, \textit{1609} (black solid lines) compared to summary domains (gray areas) obtained from synthetic datasets simulated under $50$ kept ABC posterior samples (re-sampled using the corresponding weights). The red dashed lines represent summaries obtained under the corresponding ABC posterior means.}
		\label{fig:fittedsummaries_ratdata}
	\end{centering}
\end{figure}

Fitting real data, we do not have ground truth parameters to compare our ABC results with. However, we check here whether the obtained results fit the model, which would advocate in favour of both the model and the inferential procedure.
Figure \ref{fig:fittedsummaries_ratdata} reports the estimated structure-based summary statistics \eqref{s} from the four investigated real action potential recordings (black solid lines) in comparison with the summaries estimated from synthetic datasets simulated under fifty parameter values sampled from the ABC posterior according to their weights (gray areas) and under the ABC posterior means reported in Table \ref{table:ABCresults_ratdata} (red dashed lines).

We observe a good fit of the summary statistics from real and synthetic data in all considered scenarios, with the gray bands for the densities being narrower than those for the spectral densities. Note also that the peaks of the real data spectral densities (resp. densities) for the \lq\lq stimulating\rq\rq \ scenarios are higher (resp. lower) than those for the \lq\lq resting\rq\rq\  scenarios, both features being captured by the summaries coming from the fitted model. 

\newpage

\begin{figure}[H]
	\begin{centering}
		\includegraphics[width=0.92\textwidth]
		{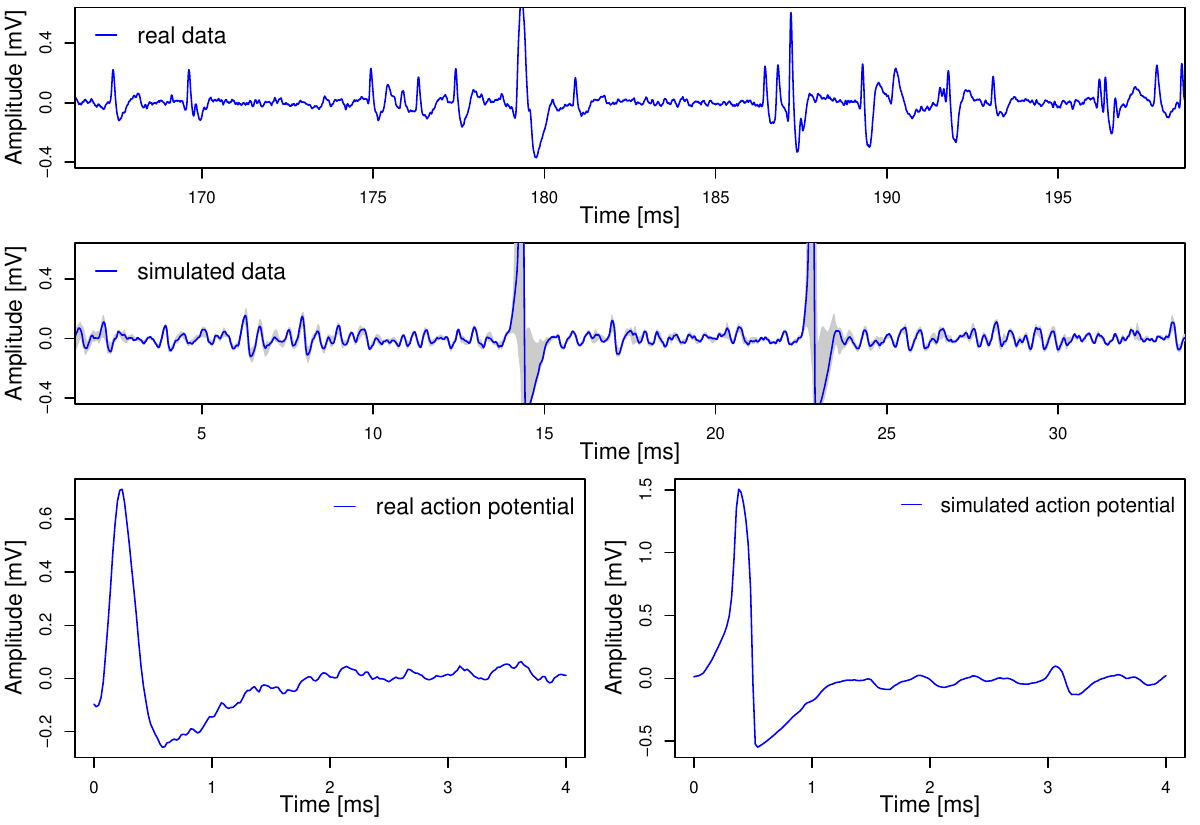}
		\caption{Exemplary 30 ms data snippet from recording \textit{1553} (top panel) compared to a simulated dataset obtained under the corresponding ABC posterior means (middle panel, blue solid line) and to the posterior predictive band (middle panel, shaded gray area) based on $500$ synthetic datasets simulated using the same seed for pseudo-random number generation and different parameter values re-sampled from the ABC posterior distribution. Exemplary real action potential (bottom left panel) taken from recording \textit{1553} compared to a simulated action potential (bottom right panel) taken from the blue  synthetic dataset visualised in the middle panel.}
		\label{fig:fittedpaths_ratdata}
	\end{centering}
\end{figure}

This nice fit is also observed in Figure \ref{fig:fittedpaths_ratdata}. There, we compare 
an exemplary $30$ ms snippet from the real action potential recording \textit{1553} (top panel) with a $30$ ms simulated dataset generated under the corresponding posterior means reported in Table \ref{table:ABCresults_ratdata} (middle panel, blue solid line) and a $30$ ms posterior predictive band (middle panel, shaded grey area) of synthetic datasets simulated under 500 kept ABC posterior samples (re-sampled using the corresponding weights). Moreover, we compare an action potential taken from the real data recording \textit{1553} (bottom left panel) 
with another taken from the reported simulated dataset based on the ABC posterior means (bottom right panel). The synthetic action potential closely resembles the shape of the real action potential, including the depolarization (rapid increase of voltage), repolarization (rapid decrease of voltage) and refractory (time during which the cell cannot repeat an action potential) periods. 
Not all kept posterior samples yield paths with action potentials arising at the same times as those obtained when simulating from the ABC posterior means (blue line in the middle panel of Figure \ref{fig:fittedpaths_ratdata}), which explains the slightly broader gray band around the spiking times. However, those paths contain at least one of the two exemplified action potentials and/or show spikes at other epochs, so the whole ABC posterior distribution reproduces the desired neural activity, with posterior predictive densities and spectral densities nicely matching those derived from the real dataset (cf. Figure \ref{fig:fittedsummaries_ratdata}).


\section{Conclusion}
\label{sec:6_FHN}

We propose a SMC variant of the SBP ABC method initially presented in \cite{Buckwar2019} for a different class of SDEs, with the goal of estimating the parameters of the stochastic FHN model. This is a well known SDE used to describe the spiking activity of a single neuron at the intracellular level. The proposed SBP SMC-ABC method relies on summary statistics based on the ergodicity property of the model \cite{Leon2018} and on a recently proposed structure-preserving numerical splitting scheme for reliable and efficient synthetic data generation \cite{Buckwar2022}.

While existing estimation methods often require complete observations, a non-degenerate noise structure, globally Lipschitz coefficients or a very complicated architecture, our method overcomes the associated difficulties, is easy-to-grasp and straight-forward to implement and to parallelise. Moreover, in contrast to other inference procedures (which, e.g. require to fix the time scale separation parameter $\epsilon$, see \cite{Ditlevsen2019}), the SMC SBP-ABC algorithm succeeds in simultaneously estimating all four model parameters of the stochastic FHN model, even for fewer observations and relatively small observation time horizons.  

While there exist very few real data studies using the FHN model, the presented analysis of real action potential is the first for the considered stochastic version of the model. Here, we investigate real data
recorded under different experimental conditions (resting and cutaneous stimulation) in an adult female rat \cite{METCALFE2020}. The inferred posterior distributions are in agreement with the different experimental settings and lead to satisfactory goodness-of-fit plots, suggesting the goodness of the considered model and the proposed estimation approach.

Besides successfully deriving marginal posterior densities from both synthetic and real data, we also discuss the influence of different key ingredients of ABC. For example, the proposed structure-based summaries avoid possible biases which may be instead introduced by the canonical ones, while other choices of percentiles (e.g. the 25th), distances (e.g. the Wasserstein) or priors (e.g. log-normal and exponential) have a limited impact on the results. In terms of proposal samplers, which one to choose depends on the available computational budget. Indeed, \texttt{olcm} (and other guided sequential ABC schemes proposed in \cite{PicchiniTamborrino2022}) leads to more accurate posterior inference than \texttt{standard} (the perturbation kernel considered here) for low/moderate computational budgets, with the differences becoming negligible as the computational budget increases. On the other hand though, \texttt{olcm} and the guided sequential schemes are characterised by more involved proposal means and covariances, introducing a computational overhead with respect to \texttt{standard}.

Finally, we note that the proposed method can be applied to any other ergodic SDE for which a computationally efficient and reliable numerical method for synthetic data generation is available. The exploration of more complex stochastic neuronal models via the proposed ABC methodology (or suitable adaptations thereof) is certainly of interest and will be considered in future works. Moreover, motivated by \cite{Dyer2023}, investigations using path signatures as summary statistics within ABC-inference for SDE models are currently undergoing, and may lead to successful inference also for non-ergodic models and/or multivariate observed components.





\appendix

\section{Computational cost of different summary statistics}\label{AppExtraA}

In Section \ref{sec:impact_ingredients_algorithm}, we discussed the impact of canonical and structure-based summary statistics on the quality of the inference, observing how the latter requires a lower computational budget (in terms of model simulations) than the former to achieve a desired accuracy level. In this section, we investigate the computational cost of calculating these summaries and the impact it has in terms of runtimes of the algorithms. We used the R-package \textit{microbenchmark} to measure the time it takes to compute the structure-based and canonical summary statistics for all considered datasets, repeating such operation \textit{sequentially} 5,000 times per dataset. Note that the actual evaluation time, over several cores, will be lower. The median evaluation times, in microseconds $[\mu s]$, were obtained on a MacbookPro 2020 (Processor 1.4GHz QUad-Core Intel Core i5, Memory: 8 GB 2133 MhZ) and are reported in Table \ref{Table1}.

Interestingly, evaluating the canonical summaries is $1.6-1.75\times$ faster than the structure-based ones for datasets with $n=625, 1250, 2500$ observations, but only $1.3\times$ faster for datasets with $n=5000, 10000$ observations, when comparing the median evaluation times.  In particular, computing the canonical summaries for the largest dataset takes a median time of $4.428 \ \mu\textrm{s}$ compared to $5.775 \ \mu\textrm{s}$ for the structure-based.

\begin{figure}
	\centering
	\includegraphics[width=.7\textwidth]{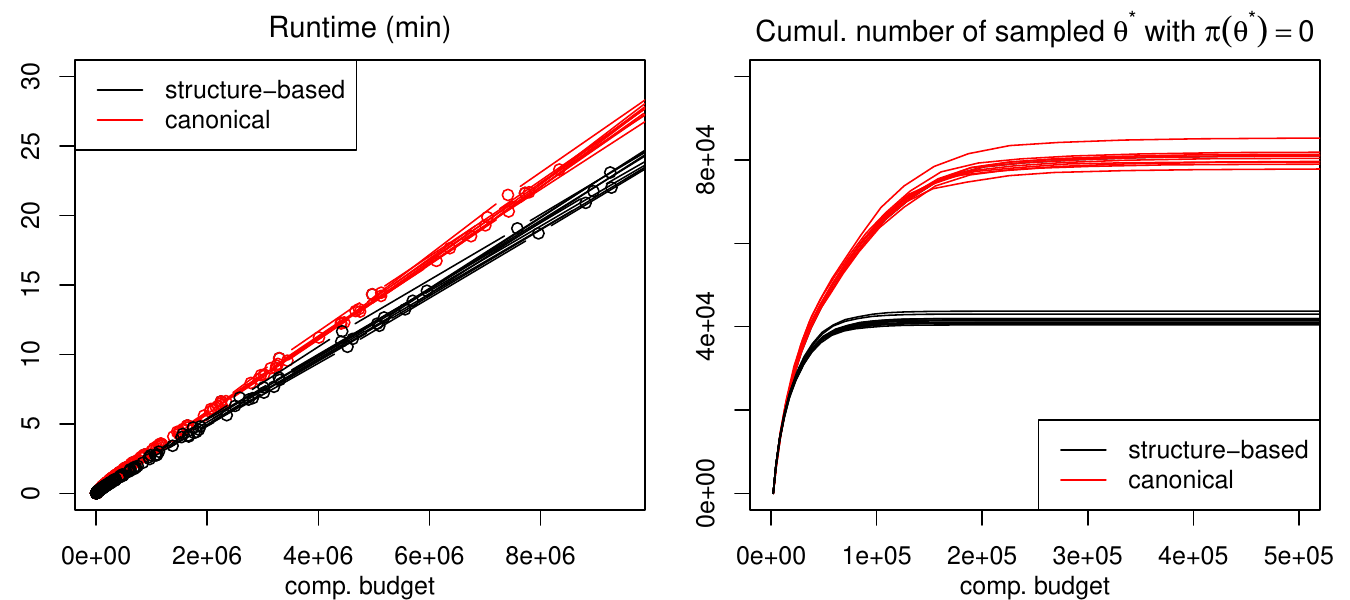}
	\caption{Runtimes in minutes (left panel) and cumulative number of immediately rejected samples $\theta^{*}$ (right panel) of ten runs of the SMC-ABC Algorithm \ref{alg:SMC_SBP_ABC} with canonical (red lines) and structure-based summaries (black lines), respectively, as functions of the computational budget, for the setting $T_\textrm{obs}=200$ and  $\Delta_\textrm{obs}=0.02$.}
	\label{runtime10m}
\end{figure}

The impact on the runtimes of the algorithm is negligible though, with the SMC-ABC algorithm with structure-based summaries being actually slightly faster than that with canonical summaries. This can be observed in Figure \ref{runtime10m} (left panel), where we report the runtimes (in minutes) of the respective algorithms as functions of the computational budget for ten runs each (launched on a machine with $48$ cores at the University of Warwick). The slightly lower runtimes obtained under the structure-based summaries may be explained by the fact that fewer samples $\theta^*$ were immediately rejected due to being outside the prior domain (cf. Line 19 of Algorithm \ref{alg:SMC_SBP_ABC}), as observed in the right panel of Figure \ref{runtime10m}. This is a further advantage of the structure-based~summaries.

\begin{table}
	\centering
	\scalebox{0.95}{
		\begin{tabular}{ccccc}
			\hline
			$T_\textrm{obs}$&$\Delta_\textrm{obs}$ & n&Summary statistics& Median evaluation time $[\mu s]$\\ \hline  \hline    200& 0.02& $10^4$& struct.-based&   5775.1880 \\
			&& &canonical &4428.1330 \\ 
			& 0.04&5000&struct.-based&3332.5880 \\
			&&&canonical &2487.1360  \\ 
			& 0.08&   2500&struct.-based&  2074.2020\\
			& &&canonical&  1297.2495  \\ \hline
			100& 0.02& 5000&struct.-based&  3316.5125 \\
			& &&canonical  & 2475.3875 \\ 
			&0.04& 2500& struct.-based& 2067.1440\\ 
			&&&canonical & 1293.1045\\ 
			&0.08&1250& struct.-based&1550.1925\\
			&& &canonical  & 965.1840 \\ \hline
			50& 0.02& 2500&struct.-based&  2059.7680\\
			&&&canonical & 1289.8485\\ 
			&0.04& 2500& struct.-based& 1546.2915\\
			&&&canonical  & 968.1145\\ 
			& 0.08& 625&struct.-based&     1403.1120 \\
			&&&canonical& 802.2475 \\ \hline
	\end{tabular}}
	\caption{Median times (in microseconds $\mu s$) computed over 5000 sequential repetitions to evaluate the structure-based and canonical summary statistics for different datasets.}
	\label{Table1}
\end{table}

\newpage

\section{Impact of the prior distribution}\label{AppendixPrior}

In this section, we investigate the proposed SBP SMC-ABC method (i.e., Algorithm \ref{alg:SMC_SBP_ABC} with structure-preserving splitting simulation, structure-based summaries and \texttt{standard} Gaussian proposal sampler) with different prior distributions for the model parameters. In particular, we fix $T_\textrm{obs}=200$ and $\Delta_\textrm{obs}=0.02$, and compare the estimation results based on the uniform priors \eqref{eq:initial_prior_FHN} (cf. black lines in Figure \ref{fig:ABC_CIs} and Figure \ref{fig:ABC_posteriors} of Section \ref{SectionEqui}) with those obtained from log-normal and exponential priors, given by
\begin{eqnarray*}
	&&	\epsilon \sim \textrm{LogN}(0,1), \quad \gamma \sim \textrm{LogN}\Bigl(0,\frac{1}{2}\Bigr), 
	\quad \beta \sim \textrm{LogN}(0,1), \quad \sigma \sim \textrm{LogN}\Bigl(0,\frac{3}{4}\Bigr),
	\\
	&&	\epsilon \sim \textrm{Exp}(3), \quad \gamma \sim \textrm{Exp}\Bigl(\frac{1}{2}\Bigr),
	\quad \beta \sim \textrm{Exp}\Bigl(\frac{1}{2}\Bigr), \quad \sigma \sim \textrm{Exp}(1),
\end{eqnarray*}
respectively. Here, $\textrm{LogN}(\mu,\sigma)$ denotes a logNormal distribution where the underlying Normal has mean $\mu$ and standard deviation $\sigma$, while $\textrm{Exp}(\lambda)$ is an exponential with rate parameter $\lambda$.
All investigated priors have means, medians and modes different from the true values. Moreover, they all account for the positivity of the model parameters, the log-normal and exponential ones being defined on the entire parameter space $(0,\infty)^4$. 

\begin{figure}
	\centering   \includegraphics[width=1.0\textwidth]{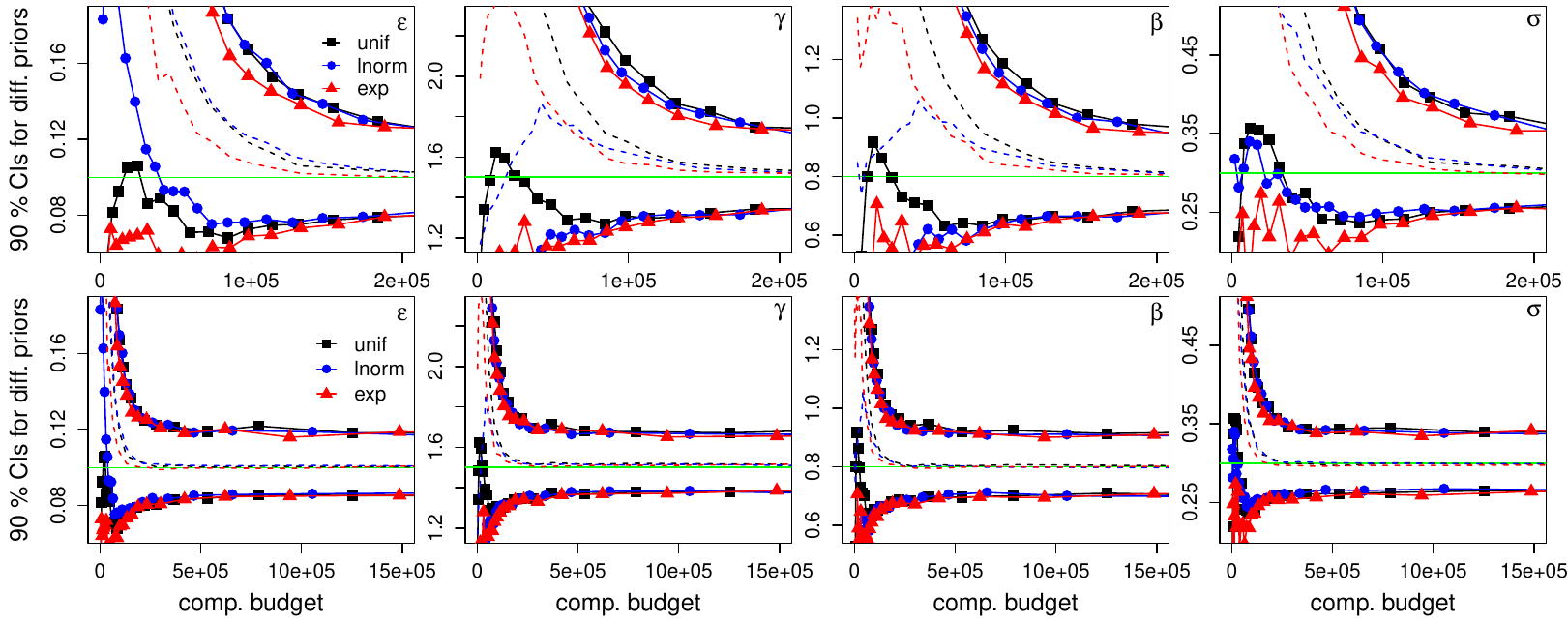}
	\caption{$90$\% CIs of the marginal posterior densities of the parameter vector $\theta$ \eqref{eq:theta} of the stochastic FHN model \eqref{FHN} obtained from Algorithm \ref{alg:SMC_SBP_ABC} with uniform (black squares), log-normal (blue circles) or exponential (red triangles) priors. The green horizontal lines and the dashed lines indicate the true parameter values and the 
		corresponding posterior medians,~respectively.}
	\label{fig:CIs_diffPriors}
\end{figure}

\begin{figure}
	\begin{centering}
		\includegraphics[width=1.0\textwidth]
		{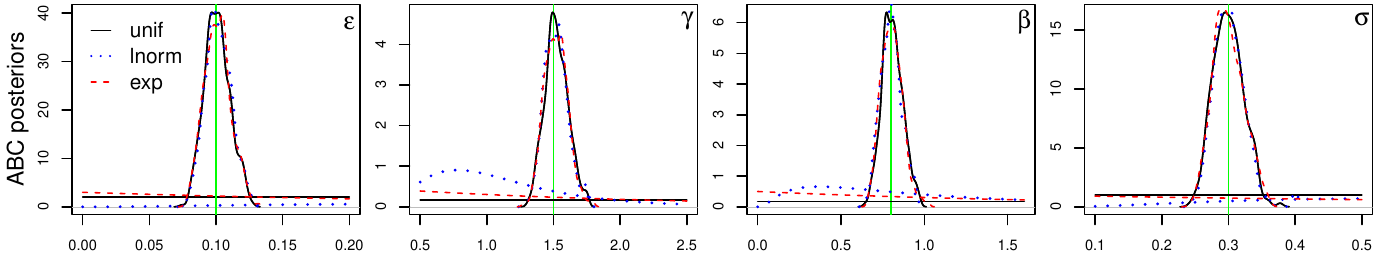}
		\caption{Marginal posterior and prior densities of the parameter vector $\theta$~\eqref{eq:theta} of the stochastic FHN model~\eqref{FHN} obtained from Algorithm \ref{alg:SMC_SBP_ABC} with uniform (black solid lines), log-normal (blue dotted lines) or exponential (red dashed lines) priors, when using a computational budget of $\textrm{Nsim}_\textrm{max}=10^6$. The green vertical lines indicate the true parameter values.}
		\label{fig:Posteriors_diffPriors}
	\end{centering}
\end{figure}

As expected, the prior distribution impacts the inferential results in the first few iterations of the algorithm,  when the used computational budget is small. However, as the algorithm evolves, the impact of the chosen prior distribution on the inferential results becomes negligible. This can be observed in Figure \ref{fig:CIs_diffPriors}, where we report the $90$\% CIs of the derived ABC posterior densities for $\theta$ \eqref{eq:theta} under uniform (black squares),
log-normal (blue circles)
and exponential (red triangles)
priors as a function of the computational budget. The CIs obtained under the different priors vary initially (see the top panels) up to a computational budget of approximately $2\cdot 10^5$, when they overlap. In Figure \ref{fig:Posteriors_diffPriors}, we report the ABC posterior densities obtained for a computational budget of $\textrm{Nsim}_\textrm{max}=10^6$ (black solid, blue dotted and red dashed lines for uniform, log-normal and exponential priors, respectively), which hardly differ from each other, suggesting a marginal impact of the chosen prior on the inferential results.

\section{Impact of the proposal kernel}\label{AppendixA}

In this section, we first define the Gaussian \texttt{olcm} proposal sampler introduced in \cite{Filippi2013} and then compare its performance with the Gaussian \texttt{standard} proposal sampler used throughout the paper (and also proposed in \cite{Filippi2013}). Alternative sequential algorithms, such as SIS-ABC and several guided sequential ABC schemes (both SIS-ABC and SMC-ABC), recently proposed in \cite{PicchiniTamborrino2022}, have also been considered. 
As their performance is comparable to that of the SMC-ABC algorithm with the \texttt{olcm} sampler (and better when using canonical summaries, which however we discourage to choose), we do not report them here.

\vspace{-0.2cm}
\subsection{Gaussian optimal local covariance matrix (\texttt{olcm}) sampler}

The Gaussian \texttt{standard} proposal sampler $K_r(\cdot|\theta_j)=\mathcal{N}(\theta_j,2\hat \Sigma_{r-1})$ considered throughout the paper is characterised by a mean $\theta_j$ sampled from the weighted set $\{(\theta_{r-1}^{(j)},\omega_{r-1}^{(j)})_{j=1}^N\}$ and a covariance matrix which is not specific for $\theta_j$, as it is computed by using all kept particles at the previous iteration (see Section \ref{SecpropSampler}). The Gaussian \texttt{olcm} proposal sampler instead, is a perturbation kernel with a covariance matrix specific for the sampled $\theta_j$. Among the $N$ accepted particles at iteration $r-1$, consider the subset of $N_0\leq N$ weighted particles whose corresponding distances (between the simulated and the reference data) are also smaller than $\delta_r$, i.e.
\begin{equation*}
	\{\tilde{\theta}_{r-1,l},\mu_{r-1,l}\}_{1\leq l\leq N_0}=\biggl\{\biggl({\theta}_{r-1}^{(j)}, \frac{w_{r-1}^{(i)}}{\bar{\mu}_{r-1}} \biggr), \text{ such that } d(s(y),s(y_{\theta^{(j)}_{r-1}})<\delta_r, \quad j=1,...,N \biggr\}
	,\label{eq:N0-particles}
\end{equation*}
where $\bar{\mu}_{r-1}$ is a normalisation constant such that $\sum_{l=1}^{N_0}\mu_{r-1,l}=1$. As we choose the new threshold level $\delta_r$ as the $p$th percentile of the $N$ accepted distances at iteration $r-1$ (see Section~\ref{secthreshold}), we already know that the set above will contain the corresponding $N_0=p\% N$ particles. Then, at iteration $r$, the \texttt{olcm} proposal is given by 
$K^\textrm{olcm}_r(\cdot|\theta_j)=\mathcal{N}(\theta_j,\Sigma^{\mathrm{olcm}}_{\theta_j})$, 
where
$
\Sigma^{\mathrm{olcm}}_{\theta_j} = \sum_{l=1}^{N_0}\mu_{t-1,l}(\tilde{\theta}_{r-1,l}-\theta_j)(\tilde{\theta}_{r-1,l}-\theta_j)^\top$.  Accepted particles are given unnormalized weights $\tilde{w}^{(j)}_r$ as in Line 25 of Algorithm \ref{alg:SMC_SBP_ABC}, with $k_r(\cdot|\theta_{r-1}^{(l)})$ being now the density of $\mathcal{N}(\theta_{r-1}^{(l)},\Sigma^\textrm{olcm}_{\theta_{r-1}^{(l)}})$. As the constructed covariance matrix $\Sigma^{\mathrm{olcm}}_{\theta_j}$ is now specific for each particle $\theta_j$, it will have to be computed for every $j$th particle, $j=1,\ldots, N$,  introducing thus a small but non-negligible overhead in the~computations. 

\vspace{-0.2cm}
\subsection{Inference under different proposal samplers}
\vspace{-0.1cm}

\begin{figure}
	\centering
	\includegraphics[width=1\textwidth]{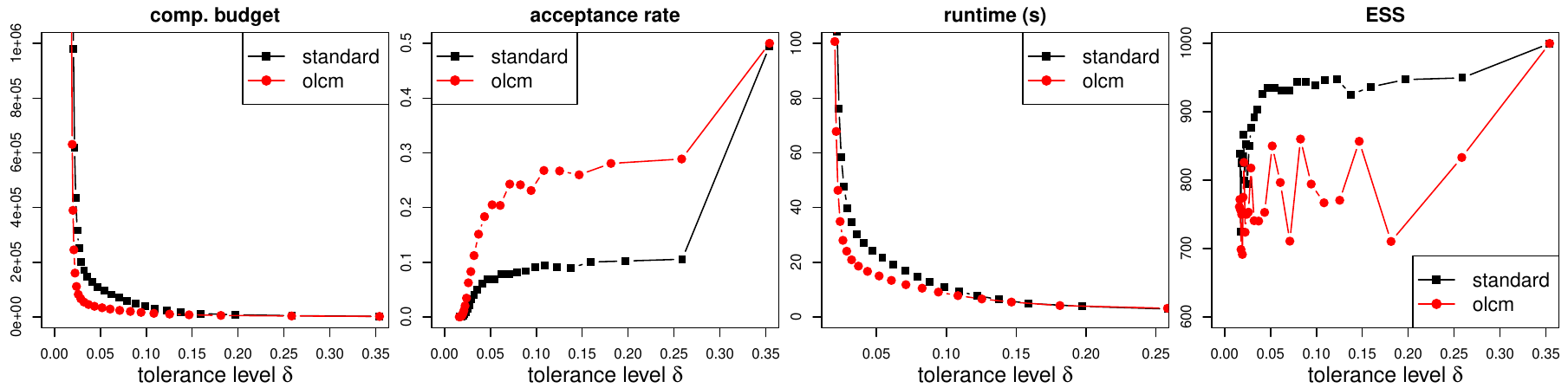}
	\caption{Computational budget, acceptance rate, runtime (in seconds) and ESS as a function of the tolerance level $\delta$ when using the SBP SMC-ABC algorithm with either the \texttt{standard} (black lines) or the \texttt{olcm} (red lines) sampler. 
	}
	\label{FeatureModelBased}
\end{figure}

In this section, we investigate the impact of the Gaussian \texttt{standard} and \texttt{olcm} proposal samplers on the inference results, considering not only the obtained marginal ABC posterior densities, but also other quantities of interest. In particular, in Figure \ref{FeatureModelBased}, we report  the computational budget, acceptance rates (estimated as $N$ divided by the number of sampled particles at iteration $r$), the ESS (approximated at iteration $r$ as $1/(\sum_{j=1}^N (w_r^{(j)})^2)$) and the algorithm runtimes as a function of the tolerance level $\delta$. Remember that the first iteration of the algorithm consists of acceptance-rejection ABC, so the perturbation kernel does not play a role there. Moreover, as the threshold levels $\delta$ are automatically decreased across iterations (with  $\delta_1\approx 0.35$), the algorithm is \lq\lq evolving\rq\rq\ (in terms of successive iterations) towards smaller values of $\delta$, so from right to left in the plots of Figure \ref{FeatureModelBased}. By looking at the first panel, we can see that \texttt{olcm} needs fewer model simulations than \texttt{standard} to achieve a certain tolerance level, which is consistent with the higher acceptance rates (second panel) across all algorithm iterations, as soon as the proposal samplers are used, i.e. from the second iteration. For example, \texttt{olcm} and \texttt{standard} achieve an approximate threshold level of 0.024 and 0.047, respectively, for a computational budget of $10^5$ model simulations, requiring approximately 390000 and 1562000 model simulations to achieve a threshold around 0.02 in 100 $s$ and 256 $s$, respectively. This is also consistent with the lower runtimes of \texttt{olcm} than \texttt{standard} (third panel). When looking at the ESS (fourth panel), a number in $[1,N]$ measuring the effectiveness of the sampler, i.e., how many particles are \lq\lq relevant\rq\rq\ at each iteration, so the higher the better, we can see that, while high for both perturbation kernels,  \texttt{standard} has higher ESS than \texttt{olcm} throughout the entire run of the algorithm. 

\begin{figure}
	\centering   \includegraphics[width=1.0\textwidth]{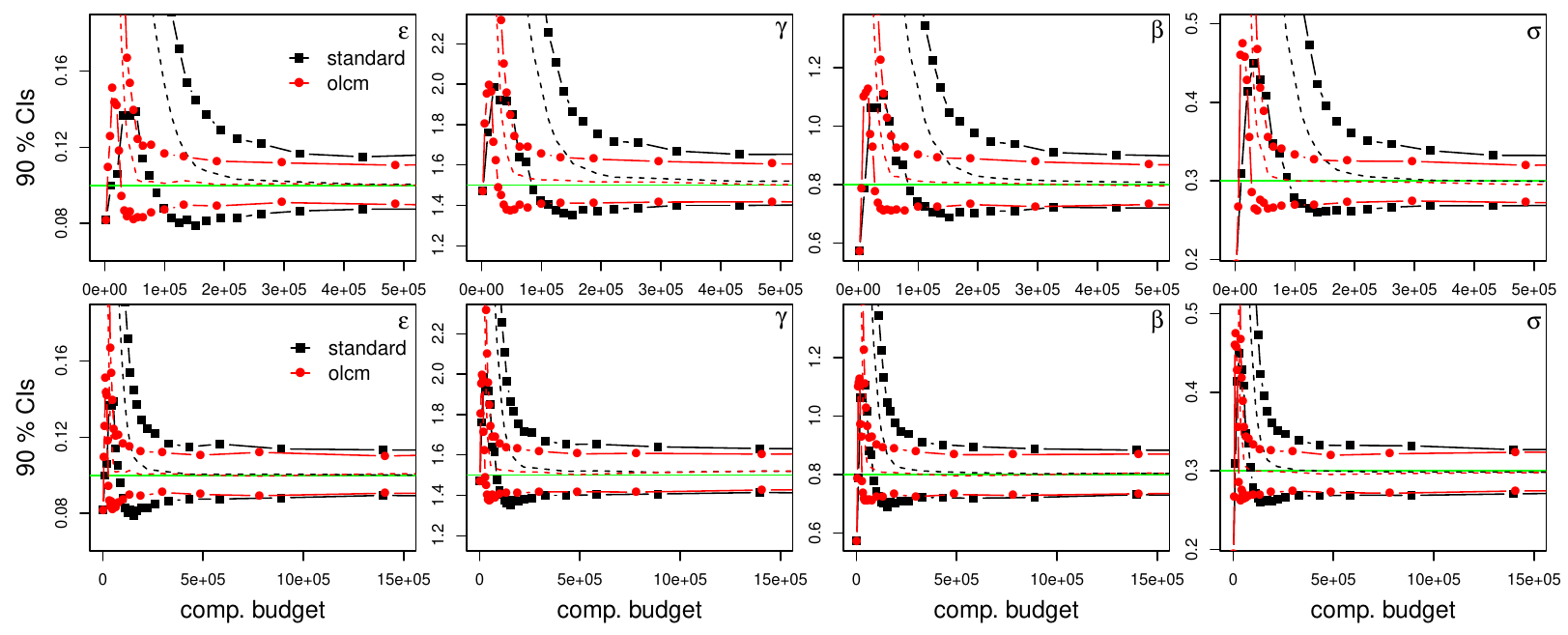}
	\caption{90\% CIs of the marginal posterior densities of $\theta$ \eqref{eq:theta} of the stochastic FHN model \eqref{FHN} obtained from the SBP SMC-ABC algorithm with the \texttt{standard} 
		(black lines) and \texttt{olcm} (red lines) samplers as a function of the computational budget. The green horizontal lines and the dashed lines indicate the true parameter values and the corresponding posterior medians,~respectively.}
	\label{CIsModelBased}
\end{figure}

\begin{figure}
	\centering
	\includegraphics[width=1.0\textwidth]{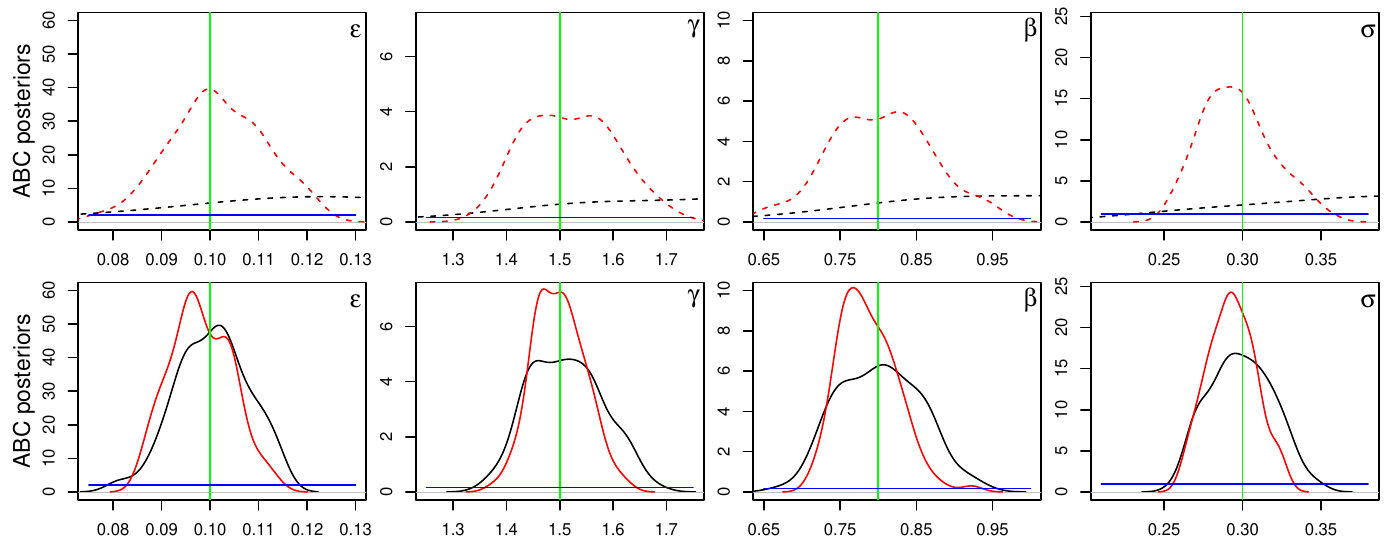}
	\caption{Marginal posterior densities of the parameter vector $\theta$ \eqref{eq:theta} of the stochastic FHN model~\eqref{FHN} obtained from the SBP SMC-ABC algorithm with \texttt{standard} (black lines) and \texttt{olcm} (red lines) when using a computational budget of approximately $10^5$  (dashed lines, top panels) and $\text{$9$-$13$}$~million (solid lines, bottom panels) model simulations, respectively. The green vertical lines and the blue horizontal lines indicate the true parameter~values and the prior densities,~respectively.}
	\label{CIsModelBasedfinal}
\end{figure}

While these quantities are important and of interest, what matters the most is obviously the quality of the inference. In Figure \ref{CIsModelBased}, we report the 90\% CIs (between the solid lines) and the median (dashed lines) of the marginal posterior densities obtained using the \texttt{standard} (black lines) and the \texttt{olcm} (red lines) proposal samplers, as a function of the computational budget. 
The choice of the sampler influences the inferential results, especially for a low/moderate computational budget. 
Indeed, \texttt{olcm} requires many fewer model simulations than \texttt{standard} to obtain narrow CIs which are centered around the true parameters, see the top panels in Figure \ref{CIsModelBased}. When the computational budget increases, such differences, while still present, become negligible. This can also be observed in Figure \ref{CIsModelBasedfinal}, where we report the SBP SMC-ABC marginal posteriors (black lines for \texttt{standard} and red lines for \texttt{olcm}) obtained with a computational budget of  $10^5$ (top panels, dashed lines) and 9-13 millions (bottom panels, solid lines) model simulations.


\section{Real data study: ABC posterior scatter plots}\label{AppendixScatter}

In this section, we provide a more comprehensive picture of the ABC posterior distributions obtained from the four investigated action potentials \textit{1553}, \textit{1608} (animal at rest) and \textit{1554}, \textit{1609} (animal stimulated), cf. Section \ref{sec:5_FHN}.

\begin{figure}[H]
	\centering
	\includegraphics[width=1.0\textwidth]{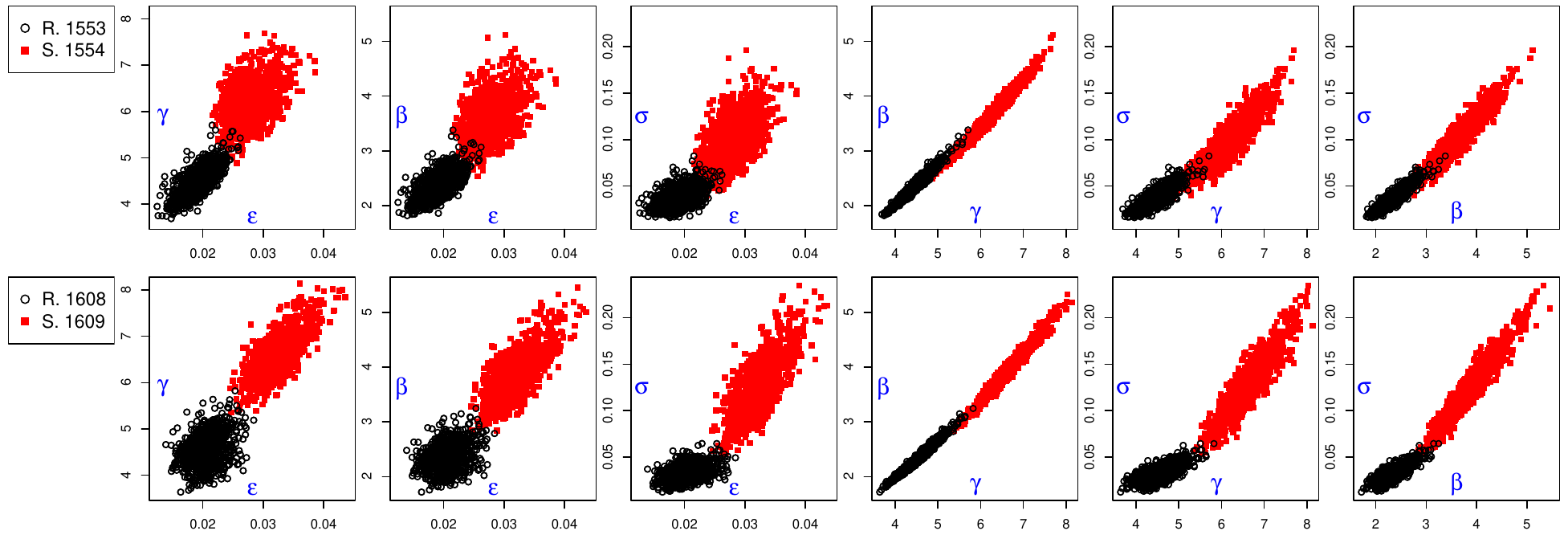}
	\caption{Scatter plots of pairs of the kept ABC posterior samples obtained from the real action potential datasets \textit{1553} (animal at rest, black dots in the top panels), \textit{1554} (animal stimulated, red dots in the top panels), \textit{1608} (animal at rest, black dots in the bottom panels) and \textit{1609} (animal stimulated, red dots in the bottom panels).}
	\label{fig:scatter_ratData}
\end{figure}

Figure \ref{fig:scatter_ratData} reports scatter plots of pairs of the kept ABC posterior samples, the black and red dots referring to the ``resting'' and ``stimulating'' scenarios, respectively. The posterior point clouds of the two ``resting'' (resp. ``stimulating'') scenarios are located at similar parameter regions. Those regions differ across the two scenarios, with the kept ABC posterior samples in the ``stimulating'' case shifted towards larger parameter values. This is in agreement with what observed when looking at the  marginal posterior densities reported in Figure~\ref{fig:ABCresults_ratdata} of Section~\ref{sec:5_FHN}.

Table~\ref{table:ABCresults_ratdata_cor} reports the corresponding correlation coefficients among pairs of kept ABC posterior samples. All four action potential datasets yield ABC posterior samples with a strong correlation among the parameters $\gamma$ and $\beta$, $\gamma$ and $\sigma$, and $\beta$ and $\sigma$. The correlation among $\gamma$ and $\sigma$ and among $\beta$ and $\sigma$ is slightly larger for the two ``stimulating'' scenarios compared to the ``resting'' scenarios. In addition, we also observe differences in the pairwise parameter correlations within the two settings. For example, we obtain a rather strong correlation between $\epsilon$ and $\gamma$ for the ``resting'' scenario \textit{1553}, and a rather weak one for the ``resting'' scenario \textit{1608}. Similarly, we observe a rather strong (resp. weak) correlation between $\epsilon$ and $\gamma$ for the ``stimulating'' scenario \textit{1609} (resp. \textit{1554}). 

\begin{table}[H]
	{\small  
		\caption{ABC posterior correlation coefficients of pairs of parameters, obtained under four different real action potential datasets.}
		\vspace{-0.5cm}
		\label{table:ABCresults_ratdata_cor}
		\begin{center}
			\scalebox{0.9}{
				\hspace{-1.0cm}
				\begin{tabular}{cc}
					\hline 
					case & ABC posterior correlations \\ \hline 
					Rest. \textit{1553} & $\epsilon,\gamma$: 0.755; $\epsilon,\beta$: 0.668;  $\epsilon,\sigma$: 0.507; $\gamma,\beta$: 0.977; $\gamma,\sigma$: 0.828; $\beta,\sigma$: 0.923    \\
					Rest. \textit{1608} & $\epsilon,\gamma$: 0.330; $\epsilon,\beta$: 0.326;  $\epsilon,\sigma$: 0.443; $\gamma,\beta$: 0.985; $\gamma,\sigma$: 0.715; $\beta,\sigma$: 0.816    \\
					Stim. \textit{1554} & $\epsilon,\gamma$: 0.430; $\epsilon,\beta$: 0.417;  $\epsilon,\sigma$: 0.522; $\gamma,\beta$: 0.986; $\gamma,\sigma$: 0.887; $\beta,\sigma$: 0.943    \\
					Stim. \textit{1609} & $\epsilon,\gamma$: 0.750; $\epsilon,\beta$: 0.734;  $\epsilon,\sigma$: 0.760; $\gamma,\beta$: 0.985; $\gamma,\sigma$: 0.908; $\beta,\sigma$: 0.958    \\
					\hline
			\end{tabular}}
	\end{center}}
\end{table}


\end{document}